%% file: main.tex
\documentclass{aa}  
\usepackage{graphicx}
\usepackage{natbib,twoopt}
\usepackage{xcolor}
\usepackage{pdflscape}
\usepackage{comment}
\usepackage{ulem}
\usepackage{float}
\usepackage{cancel}

\newcommand{\io}[2]{#1{\textsc{#2}}}

\usepackage[colorlinks=true, citecolor=blue, breaklinks=true, linkcolor=blue]{hyperref}
\bibpunct{(}{)}{;}{a}{}{,}             
\makeatletter
  \newcommandtwoopt{\citeads}[3][][]{\href{http://adsabs.harvard.edu/abs/#3}%
    {\def\hyper@linkstart##1##2{}%
     \let\hyper@linkend\@empty\citealp[#1][#2]{#3}}}
  \newcommandtwoopt{\citepads}[3][][]{\href{http://adsabs.harvard.edu/abs/#3}%
    {\def\hyper@linkstart##1##2{}%
     \let\hyper@linkend\@empty\citep[#1][#2]{#3}}}
  \newcommandtwoopt{\citetads}[3][][]{\href{http://adsabs.harvard.edu/abs/#3}%
    {\def\hyper@linkstart##1##2{}%
     \let\hyper@linkend\@empty\citet[#1][#2]{#3}}}
  \newcommandtwoopt{\citeyearads}[3][][]%
    {\href{http://adsabs.harvard.edu/abs/#3}
    {\def\hyper@linkstart##1##2{}%
     \let\hyper@linkend\@empty\citeyear[#1][#2]{#3}}}
     
\makeatother

\usepackage{txfonts}
\begin{document}

   \title{Comparative Study of Two Luminous Red Novae}

   \subtitle{I. Progenitor Modeling And Dust Formation}

   \author{M. Wavasseur\inst{1,2,3}\thanks{\href{https://orcid.org/0009-0005-9552-3887}{ORCID: 0009-0005-9552-3887}} \and N. Blagorodnova\inst{1,2,3}\thanks{\href{https://orcid.org/0000-0003-0901-1606}{ORCID: 0000-0003-0901-1606}} \and M.~A. G{\'o}mez-Mu{\~n}oz\inst{1,2,3}\thanks{\href{https://orcid.org/0000-0002-3938-4211}{ORCID: 0000-0002-3938-4211}} \and O.~R. Pols\inst{4}}

    \institute{Institut de Ciències del Cosmos (ICCUB), Universitat de Barcelona (UB), c. Martí i Franquès, 1, 08028 Barcelona, Spain 
    \and Departament de Física Quàntica i Astrofísica (FQA), Universitat de Barcelona (UB), c. Martí i Franquès, 1, 08028 Barcelona, Spain 
    \and Institut d'Estudis Espacials de Catalunya (IEEC), Edifici RDIT, Campus UPC, 08860 Castelldefels (Barcelona), Spain
    \and  Department of Astrophysics/IMAPP, Radboud University, P O Box 9010, NL-6500 GL Nijmegen, The Netherlands\\
 \email{m.wavasseur@icc.ub.edu, nblago@fqa.ub.edu}}

   \date{Received XXXX; accepted XXXX}

 
  \abstract
  {}
   {Luminous red novae are astrophysical transients associated with binary interactions driven by unstable mass transfer. They are generally interpreted as arising from common-envelope evolution, during which partial envelope ejection can lead to a merger. These interactions can liberate large amounts of gas, some of which may subsequently condense into dust. This study focuses on the two luminous red novae AT2021biy and AT2021blu, aiming to constrain the masses and orbital parameters of their binary progenitors, estimate the mass ejected during the outbursts, and infer the dust mass from the infrared evolution of their remnants.}
   {We generated two grids of binary stellar evolution tracks using the binary module of the stellar evolution code MESA, constrained by pre-outburst photometry from the Hubble Space Telescope (AT2021biy) and ground-based observations (AT2021blu). We applied mass-transfer instability criteria to select progenitors capable of merging on timescales consistent with the archival data. Guided by these models, we estimated an upper limit on the total gas mass lost during the mass-transfer phase, as well as lower and upper bounds on the envelope mass that could be ejected during a common-envelope episode given the available orbital energy. These ejecta-mass estimates were compared with values inferred from light-curve models. Finally, we modeled mid-infrared NEOWISE data to derive dust masses up to $\sim3$ years post outburst, providing an additional lower limit on the total ejecta mass.}
   {We constrained the donor masses to $M_{d}=18-23M_{\odot}$ for AT2021biy and $M_{d}=14\pm0.5M_{\odot}$ for AT2021blu. The lower-limit estimates for the envelope mass ejected at the onset of mass-transfer instability span 0.03 to 2.98 $M_{\odot}$ for AT2021biy, and 0.02 to 0.1  $M_{\odot}$ for AT2021blu. Comparison with light-curve-based models reveals a higher degree of consistency within the intermediate mass-ratio regime ($3\lesssim q\lesssim10$ for AT2021biy and $5<q\lesssim15$ for AT2021blu). The dust masses we measured are 1–5 orders of magnitude lower than the estimated ejected envelope masses, implying that only a small fraction of the ejected gas condenses into dust. The dust-mass evolution is consistent with shock-driven interactions, indicating the presence of pre-existing circumstellar material, in line with the significant pre-outburst mass loss predicted by our MESA models.}
   {}
   
   \keywords{binaries:general, common envelope evolution—stars:winds,outflows,dust formation—stars:individual:AT2021biy—stars:individual:AT2021blu}

   \maketitle
   \nolinenumbers
%

\section{Introduction}

Binary systems statistically represent the most likely environment for massive stars, where more than two-thirds are expected to live in such systems \citep{Sana12,2013A&A...550A.107S,2014ApJS..215...15S,2022RAA....22b5009G} and a significant fraction of these binaries have short initial orbital periods of less than 10 days \citep[e.g.,][]{Moe2017ApJS,2022A&A...658A..69B}. Moreover, studies show that among massive main-sequence (MS) stars, 30$\%$ have already undergone a phase of mass transfer through interaction with a companion, and $\sim$10$\%$ have merged \citep{2014ApJ...782....7D}. Given the high incidence of binary interactions, it is therefore crucial to study these systems in order to better understand stellar evolution. In particular, the common-envelope (CE) phase \citep[][]{1976IAUS...73...75P} plays a decisive role in determining the final fate of interacting binaries.

The CE phase is crucial for explaining the formation of compact binaries such as cataclysmic variables, close double white dwarfs and Type Ia supernova progenitors \citep[see, e.g.,][]{1979A&A....78..167M,1993ApJ...406L..15I,2014ARA&A..52..107M} and gravitational wave sources \citep{2000A&A...360.1011N,2016Natur.534..512B,VignaGomez2020}, such as binary neutron-star or black-hole (BH) mergers, hundreds of which have been detected by the LIGO-Virgo-KAGRA collaboration \citep[e.g.,][]{2016PhRvL.116f1102A,2017ApJ...848L..12A,2022arXiv221115693G}. Although alternative channels for the formation of BH mergers have been proposed, including stable mass transfer \citep[e.g.][]{2017MNRAS.471.4256V}, chemically homogeneous evolution in close, rapidly rotating binaries \citep[e.g.][]{2016MNRAS.458.2634M}, and dynamical interactions in dense stellar environments like nuclear star clusters \citep[e.g.][]{2024PhRvD.110d3023K}, the CE phase remains the canonical formation pathway. The orbital energy lost during CE significantly contributes to shrinking the binary orbital separation, explaining the observed short-period orbits.

The transition to the CE phase is triggered by unstable mass transfer, which lies at the core of the system’s runaway evolution, and primarily arises: (i) due to the donor’s structural response to mass loss \citep[dynamical instability;][]{1992PASP..104..717P,1997A&A...327..620S}, (ii) when the accretor is unable to accept the incoming mass-transfer rate, forcing mass to be ejected from the system and driving runaway angular-momentum loss, or (iii) when the donor fills its outer lobe and mass loss through the outer Lagrange point L$_{2}$ carries away sufficient angular momentum \citep[e.g.,][]{2017MNRAS.471.3200M}. The CE phase can also be triggered by tidal instability if the donor’s spin angular momentum becomes too large compared to the orbital angular momentum \citep[known as the Darwin instability;][]{1879Obs.....3...79D}. Once triggered, these instabilities may cause the donor star to expand until its outer layers engulf the companion, resulting in two stellar cores orbiting each other within the same envelope. This marks the onset of the CE phase. Once the gas ceases to co-rotate with the orbit, frictional forces arise \citep[see,][]{2013A&ARv..21...59I}, causing the orbital separation to shrink. As the companion spirals inward on a relatively short timescale, its orbital energy and angular momentum are deposited into the envelope. This process may ultimately lead either to a stellar merger, if the envelope is not successfully ejected, or to a surviving close binary, if the envelope is fully expelled.

The energy released during this inspiral and envelope interaction, temporarily increases the system’s luminosity and is reflected in the observed properties of the resulting transient event\footnote{In the remainder of this study, we use the terms \textit{transient} and \textit{event} interchangeably to refer to AT2021biy and AT2021blu.}, known as a luminous red nova (LRN). This event is particularly linked to CE evolution when envelope ejection is incomplete and the system ultimately ends in a merger. LRNe are energetic transients with typical peak luminosities in the range of $10^{38}-10^{41}\rm{ergs}^{-1}$, placing them between classical novae and supernovae in terms of brightness \citep{2007Natur.447..458K,2013A&ARv..21...59I,2015A&A...578L..10K,Pastorello2019a,Blagorodnova2021}. Their light curve usually features a relatively short-lived first peak lasting a few days, followed by a longer second peak or plateau that can persist for up to nearly a year. Their spectral energy distribution (SED) progressively shifts toward redder colors, and their spectra are characterized by narrow, prominent hydrogen emission lines, with molecular absorption features emerging in the later stages.

In the literature, we typically distinguish three main stages connecting LRNe observations to the evolution of a binary system undergoing a CE phase \citep[e.g.,][]{2020MNRAS.496.5503B}: (i) the precursor, (ii) the dynamical phase or outburst, and (iii) the remnant. The precursor corresponds to the interval during which the system undergoes Roche-lobe overflow, potentially evolving from stable mass transfer toward a highly non-conservative and unstable mass-transfer regime. During the dynamical phase, the companion is engulfed by the donor’s envelope, initiating the CE and driving rapid orbital decay. The post-outburst evolution then corresponds to the remnant phase, dominated by ejecta cooling, shock-driven interactions \citep[e.g.,][]{2017MNRAS.471.3200M}, and recombination-powered emission \citep[e.g.,][]{Matsumoto2022,2025ApJ...978...56M}, that likely power the second luminosity peak observed in LRNe. 

Only a few observational studies have attempted to constrain LRN progenitors through observations of their precursor phase, either using 1D single-star models \citep{2023A&A...671A.158P}, binary population and spectral synthesis (BPASS) model grids \citep{Cai2022AA} or binary models \citep{Blagorodnova2021}. Complementing this, theoretical studies have investigated the precursor phase using three-dimensional simulations \citep[see, e.g.][]{2017ApJ...850...59P, MacLeod2018a,MacLeod2018b,2020ApJ...895...29M, 2020ApJ...893..106M}. Pre-outburst binary interactions are expected to drive a slow, dense outflow that provides ideal conditions for dust formation \citep[see, e.g.][]{Pejcha2016a,Pejcha2016b, Pejcha2016c, 2022ApJ...937...96M}. The newly formed dust absorbs the incident radiation and reprocesses it to longer wavelengths, resulting in both optical obscuration and a shift of the emergent emission toward the infrared. The dynamical phase is characterized by relatively low outflow velocities, typically below 1000 $\mathrm{km\,s^{-1}}$, which allow the ejected gas to cool efficiently—primarily through adiabatic expansion—thereby enabling dust formation \citep[e.g.][]{Gonzalez-Bolivar2024}. Other studies have simulated this mechanism in three dimensions during the remnant phase, showing that the resulting dust can contribute significantly to the cosmic dust budget \citep{2020MNRAS.497.3166I, Bermudez2024,2026ApJ...999...16K}. 

The two LRNe AT2021biy and AT2021blu were discovered by the Asteroid Terrestrial-impact Last Alert System \citep[ATLAS;][]{2018ApJ...867..105T,2018PASP..130f4505T, 2020PASP..132h5002S} in January and February 2021, respectively. AT2021biy was identified in the nearby edge-on spiral galaxy NGC 4631, while AT2021blu was located in UGC 5829. These LRNe have been studied in \citep{Cai2022AA, 2023A&A...671A.158P, 2023ApJ...948..137K, 2026ApJ...999...16K} and also in a companion paper presenting the detailed comparative spectroscopic analysis of the two transients (hereafter Paper II; Wavasseur et al., in prep.).

To date, only a limited number of studies have proposed to combine observational constraints on the progenitor, outburst, and remnant properties \citep{Cai2022AA,2023A&A...671A.158P}, or including the estimation of the dust mass present in the remnant phase \citep{2020MNRAS.496.5503B}. Furthermore, to our knowledge, no study has yet incorporated mass transfer instability into the modelling, despite its critical importance for identifying viable progenitor binary models. 

Here, we perform a unified analysis that connects the properties of the binary progenitor to the observed precursor, outburst, and remnant, while explicitly incorporating instability criteria into the modelling. More precisely, we estimate the ejecta masses of AT2021biy and AT2021blu using three complementary approaches: (i) progenitor modelling constrained by precursor observations combined with a CE energy-budget formalism; (ii) analytical light-curve models based on photometric detections during the outburst and remnant phases; and (iii) dust-mass constraints inferred from NEOWISE mid-infrared emission during the remnant phase.

In Sect. \ref{sec:obs_constrains}, we describe the observational data used to constrain our modelling. The stellar models and our selection criteria are detailed in Sect. \ref{sec:modeling}. Sect. \ref{sec:energetics} outlines the energy-budget method we used to estimate the ejecta mass, and Sect. \ref{sec:dusty_models} shows how we measured the dust mass during the remnant phase for both objects. Finally, we discuss both our methodology and results in Sect. \ref{sec:discussion}.

\section{Observational constrains} 
\label{sec:obs_constrains}

In this section, we describe how the main observational constraints from the precursor phase, namely metallicities, luminosities, and effective temperatures, were derived and used to guide the progenitor modelling.

\subsection{Metallicity, distance and colour excess}
\label{sec:metallicity}

The evolution of massive stars is highly sensitive to metallicity \citep[e.g.][]{2020A&A...638A..55K}, yet sufficiently accurate constraints on the local composition are often unavailable. To estimate the progenitor metallicity for each event, located in different environments (see \citet{Cai2022AA,2023A&A...671A.158P} and Paper II), we adopted the metallicity of their host galaxy. 

In the case of AT2021biy, the characteristics of its host galaxy, NGC 4631, are a B-band magnitude of $m_{B}=9.75$ mag\footnotemark[1] and a distance modulus $\mu=29.36 \pm 0.15$ mag\footnote{e.g. NED: \url{https://ned.ipac.caltech.edu/}}. Using the calibration of \cite{2014AJ....147..131P}, \cite{Cai2022AA} derived an oxygen abundance of $12 +\textrm{log(O/H)}=8.39\pm0.06$ dex. By scaling this value to the solar composition $Z_{\odot}=0.0142$ reported in \citet{2009ARA&A..47..481A}, this corresponds\footnote{$Z_{biy}=0.0142\times10^{12 +\textrm{log(O/H)}-8.69}$} to a metallicity of roughly half solar of $Z_{biy}\approx0.007$. 

For consistency, we adopted the same distance and colour excess for AT2021biy as in the literature \citep[see][]{Cai2022AA}, namely $d=7.46\pm0.50$ Mpc and $E(B-V)$ = $0.271\pm0.096$ mag, assuming $R_V=3.1$.

For AT2021blu, there are no spectroscopic measurements for the oxygen abundance in its host galaxy, UGC 5829. In this case, we used its B-band magnitude of $m_{B}=13.31$ mag\footnotemark[1] and a distance modulus of $\mu=29.68 \pm 0.15$\,mag \citep[see][]{2023A&A...671A.158P} to compute its absolute B-band magnitude and estimated an oxygen abundance of $12 +\textrm{log(O/H)}=8.235$ using the \citet[][Eq.~2]{2004ApJ...613..898T} calibration. This provides a metallicity of $Z_{blu}\approx0.005$. This value is comparable to that of the Small Magellanic Cloud (SMC).  

Considering the same distance and colour excess as in \citet{2023A&A...671A.158P} of $d=8.64\pm0.61$ Mpc and $E(B-V) = 0.02$, the remote location of AT2021blu within its host galaxy and the non-detection of the Na\,{\sc i} D narrow interstellar feature at the redshift of UGC 5829 suggest that there is no significant extinction due to host galaxy dust. Therefore, we assumed that the extinction is entirely due to the Galactic foreground.

\footnotetext[1]{e.g. SIMBAD: \url{http://simbad.u-strasbg.fr/simbad/}}

\subsection{Data description} \label{sec:data description}

Archival Hubble Space Telescope (HST) observations provide coverage of AT2021biy and AT2021blu during their precursor phases. The HST photometric measurements were obtained on MJD 52854.0 for AT2021biy, which is $\sim$17.5 years before its discovery date, and on MJD 58846.9, which is $\sim1$ year before the detection of AT2021blu. The HST archival photometry of the progenitors is detailed in \cite{Cai2022AA} for AT2021biy and in \cite{2023A&A...671A.158P} for AT2021blu.

\begin{figure}[h]
    \centering
    \includegraphics[width=\linewidth]{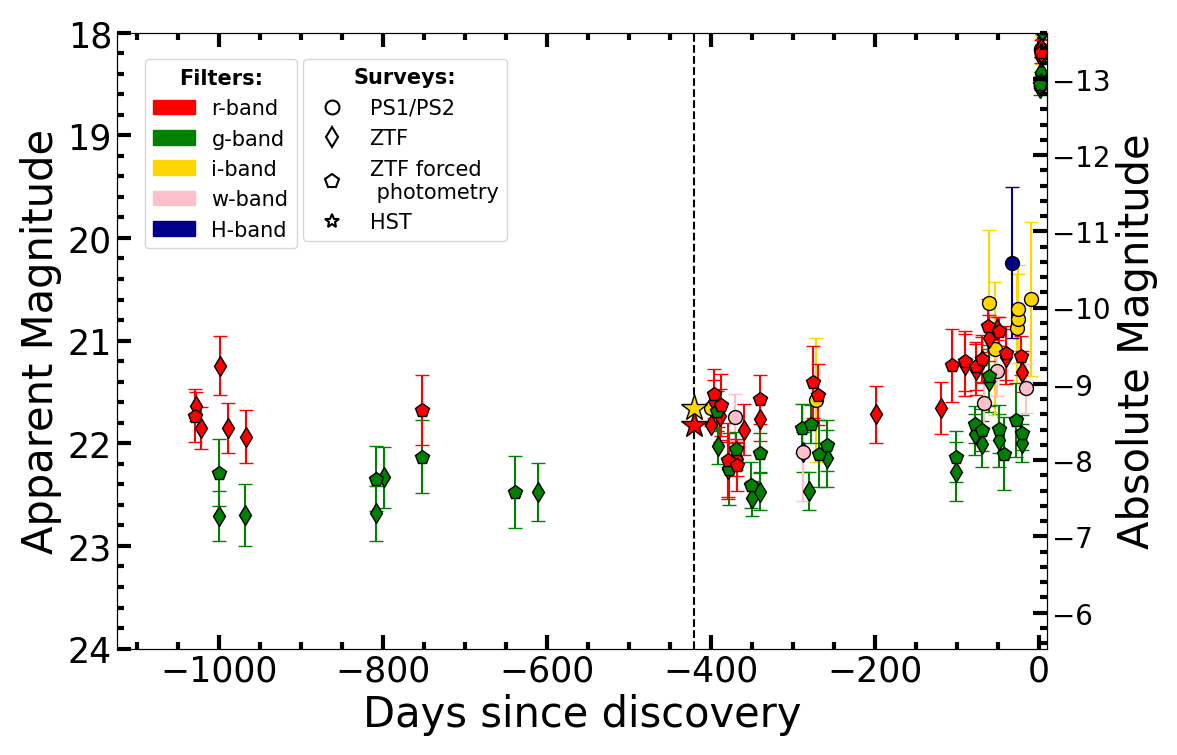}
    \caption{Multi-band light curves of AT2021blu from the ZTF and Pan-STARRS surveys, including 10-day–binned forced photometry. The single-epoch (dotted line) HST/ACS observations in the F606W and F814W filters have been converted into Sloan r and i-filters and overplotted as reference points. The reference epoch is the discovery date MJD 59246.467.}
    \label{fig:forced}
\end{figure}

In the case of AT2021blu, we complemented the HST data with forced-photometry measurements from the Zwicky Transient Facility \citep[ZTF; ][]{2019PASP..131a8002B} survey, Pan-STARRS\footnote{Panoramic Survey Telescope and Rapid Response System} photometry \citep{2016arXiv161205560C}, and the ground-based data presented in \cite{2023A&A...671A.158P}, to cover the precursor phase. More precisely, the ground-based data set consists of archival optical imaging obtained between 2006 and 2017 from several facilities (see Table 1 in \cite{2023A&A...671A.158P}), providing multi-band pre-outburst photometry at the position of AT2021blu. The 10-day–binned forced photometry extends to more than 1000 days prior to discovery and, together with the ZTF photometry and the HST observations (see Fig. \ref{fig:forced}), enables us to trace the pre-outburst activity throughout the precursor phase.

\begin{figure}[h]
    \centering
    \includegraphics[width=\linewidth]{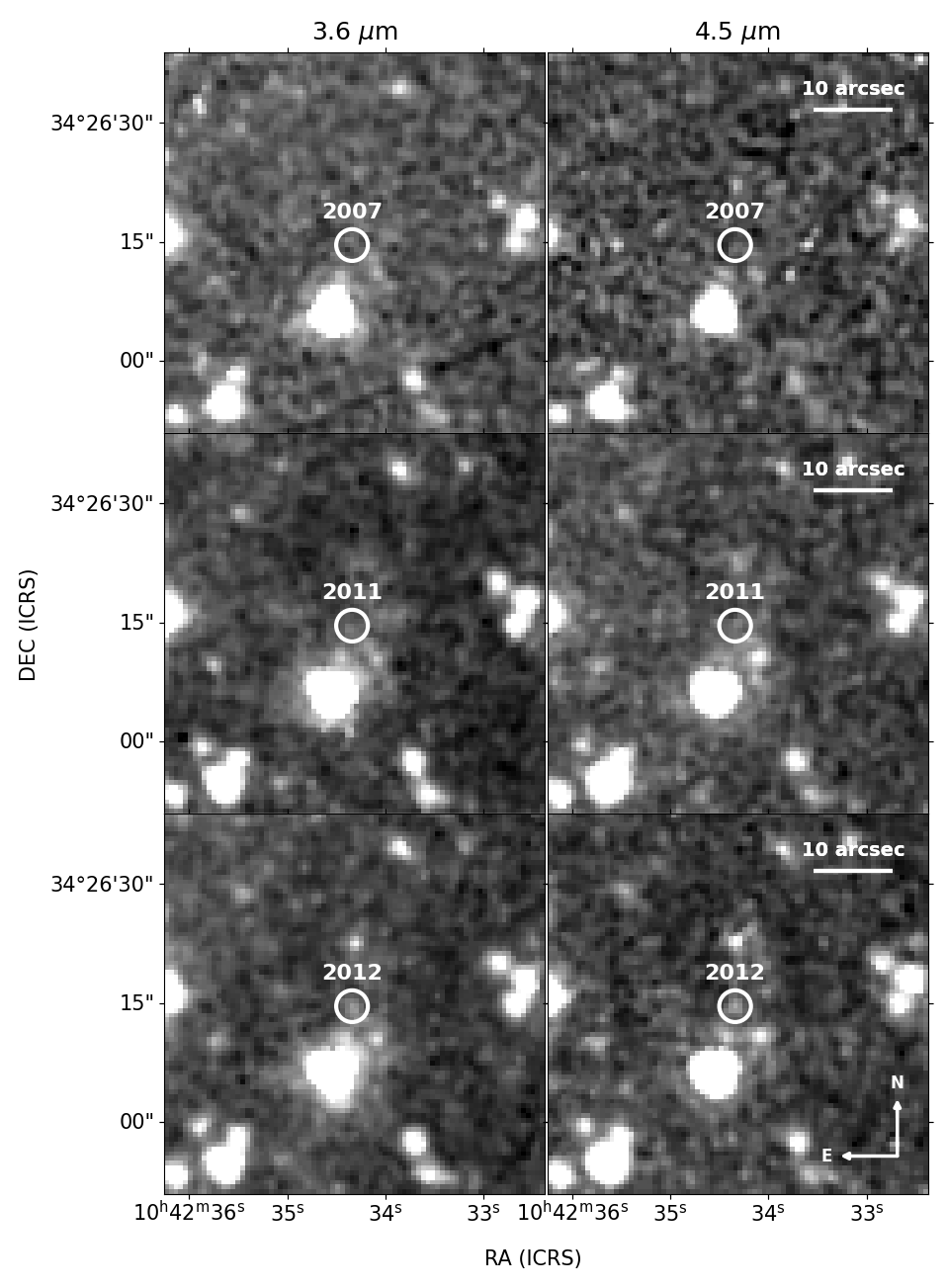}
    \caption{Spitzer/IRAC cutouts illustrating the mid-infrared photometric evolution of the transient AT2021blu from 2007 to 2012 in the 3.6$\mu m$ and 4.5$\mu m$ bands (left and right columns, respectively). Each panel shows a (48"$\times$48") field of view centered on the target.}
    \label{fig:SPITZER_cut}
\end{figure}

We also used near-IR fluxes from the NEOWISE catalogue \citep{2014ApJ...792...30M}, specifically in the W1 and W2 bands. NEOWISE observed the fields of AT2021blu and AT2021biy several times from 2010 until the end of the mission on 31 July 2024. We downloaded all the available single-exposure images from the WISE Image Service and coadded the images per season. These data provide multi-epoch constraints on the mid-IR emission and cover from $\sim80$ to $\sim1200$ days since the discovery date of each event. For AT2021blu, the mid-infrared (MIR) emission is further constrained by Spitzer photometric detections obtained under program IDs 40204 and 80025 (PIs: R. C. Kennicutt and L. van Zee, respectively), spanning the epochs December 2007, July 2011, and June 2012. These observations include series of 30 s and 100 s exposures in the 3.6$\mu m$ and 4.5$\mu m$ IRAC bands. As shown in Fig. \ref{fig:SPITZER_cut}, while there is a marginal detection in 2007, the source is clearly detected in the MIR in 2012. Using 3-pixel-radius aperture photometry and applying the standard Spitzer/IRAC aperture corrections, we measured magnitudes of $22.80\pm0.02$ and $22.12\pm0.01$ mag in the 3.6$\mu m$ and 4.5$\mu m$ bands, respectively, in 2011, and $20.40\pm0.04$ and $20.03\pm0.09$ mag in 2012. We also inspected the available Spitzer data for AT2021biy, covering the epochs 2004 and 2019, but found no convincing detection.

In summary, the data set used in this work combines new and literature data. The forced-photometry measurements from ZTF and the mid-infrared photometric detections from Spitzer are presented here for the first time, while all remaining data were taken from \citet{Cai2022AA} and \citet{2023A&A...671A.158P}.

\subsection{Photometry}
\label{sec:photometry}

In the case of AT2021biy, to estimate the fluxes in the two bands \textit{F606W} and \textit{F814W}, we performed PSF-fitting photometry of the progenitor candidate using DOLPHOT \citep{Dolphin2000, Dolphin2016}. We also added in quadrature the upper bounds of the systematic error sources associated with the zeropoint ($\sim$0.01 mag), the residual charge transfer efficiency (CTE, $\sim$0.03 mag), and the aperture correction ($\sim$0.01 mag), in agreement with the literature \citep[][]{2016AJ....152...60B}. We found \textit{F606W} $ = 22.014\pm0.034$ mag and \textit{F814W} $ = 21.411\pm0.034$ mag for AT2021biy, which is in excellent agreement with \cite{Cai2022AA}\footnote{The authors found \textit{F606W} $ = 22.038\pm0.007$ mag and \textit{F814W} $ = 21.408\pm0.008$ mag}.

We applied the same methodology in the case of AT2021blu. As described in \cite{2023A&A...671A.158P}, although two 2×380s ACS exposures were obtained in the F606W and F814W filters with dithering, AT2021blu falls outside the field of view in one of the dither positions. Consequently, with only a single flux measurement available, DOLPHOT cannot compute a repeat-frame dispersion, and the inter-exposure scatter is zero. The uncertainties related solely to the PSF and background noise therefore cannot fully represent the total error associated with the observation. To address this, we created a sample of artificial stars to quantify the empirical photometric scatter caused by crowding and blending. A large number (50,000) of fake stars of known brightness and position were injected into the science images and remeasured using the same DOLPHOT configuration as for the real data. The distribution of input–output magnitude differences in this sample, analyzed in bins of brightness and position, provides a direct estimate of the additional dispersion beyond the formal fit errors in the science images. As for AT2021biy, we also added in quadrature the upper bounds of the systematic errors. Using the adopted distance and colour excess $E(B-V)$ described in Sect.\ref{sec:metallicity}, we measured for AT2021blu \textit{F606W} $ = 21.857\pm0.034$ mag and \textit{F814W} $ = 21.262\pm0.034$ mag, which are in agreement with \cite{2023A&A...671A.158P}\footnote{The authors found \textit{F606W} $ = 21.826\pm0.008$ mag and \textit{F814W} $ = 21.226\pm0.009$}. These measurements are shown in Fig. \ref{fig:forced} together with the ZTF forced photometry and part of the ground-based detections compiled by \cite{2023A&A...671A.158P}, illustrating the photometric evolution during the precursor phase. However, the HST observations were obtained relatively shortly before the detection of the outburst. It is therefore likely that the system had already entered the dynamical phase, and we consequently did not use these observations to constrain the progenitor modelling.

Finally, we performed aperture photometry on NEOWISE images at the transient position, calibrating the fluxes using field stars with the AllWISE catalog as a reference. We then subtracted the host-galaxy contamination by taking the earliest non-detection epochs as the baseline. A total of seven and five detection epochs were found for AT2021blu (at +85, +295, +449, +659, +814, +1024, and +1180\,d MJDs) and AT2021biy (+115, +321, +480, +687, and +847\,d MJDs), respectively, between 2021 and 2024. Unfortunately, for AT2021blu, three of the seven NEOWISE detections are flagged for persistence (P) in the catalog, indicating contamination by a latent image from a bright source (in this case, a nearby RGB star) falling on the same detector pixels in a previous exposure. We therefore discard the +449, +814, and +1180 day epochs from our analysis.

\subsection{Archival SED modeling}
\label{sec: archival_data}

To derive observational constraints for our progenitor modelling, we used the archival photometric fluxes obtained during the precursor phase of both events (see Sect. \ref{sec:photometry}) to construct their SEDs. Assuming that the pre-outburst emission was dominated by the more evolved and luminous donor star, with a negligible contribution from the companion, we fitted these SEDs with single-blackbody models. From these fits, we estimated the progenitor luminosities and effective temperatures, and placed them on the Hertzsprung–Russell (HR) diagram.

We fitted the AT2021biy magnitudes presented in Sect. \ref{sec:photometry} with a single blackbody model using the Markov chain Monte Carlo (MCMC) code \texttt{BBFit}\footnote{\url{https://github.com/nblago/utils}} based on the \texttt{Python emcee} package \cite{Foreman-Mackey2013PASP}. To account for the uncertainty on the reddening, we repeated the blackbody fit three times: once using the best-estimate value of the extinction, and once at $\pm1\sigma$ around this value. This procedure yields three corresponding points in the HR diagram, which we used to delineate an uncertainty region around the derived effective temperatures and luminosities. Note that the distance to the object is also a source of uncertainty. However, its impact on the blackbody-fit coefficients is smaller than that of the reddening uncertainties (see Sect. \ref{sec:discussion}), . We obtained the following parameters for AT2021biy: $T_{\rm{eff}}=7615^{+121}_{-140}$K, $R=210^{6}_{-5}R_{\odot}$, and bolometric luminosity of $L_{bol}=1.34^{+0.03}_{-0.02}\times 10^5L_{\odot}$. Even when we account for $1\sigma$ errors in our estimates, our results remain in disagreement with those of \cite{Cai2022AA}, who found $L/L_{\odot} = 10^5L_{\odot}$ and $T_{\rm{eff}}$ = 5900K, by matching the HST observed magnitudes with synthetic photometry for a grid of single-star evolution models.

\begin{figure}
    \centering
    \includegraphics[width=\linewidth]{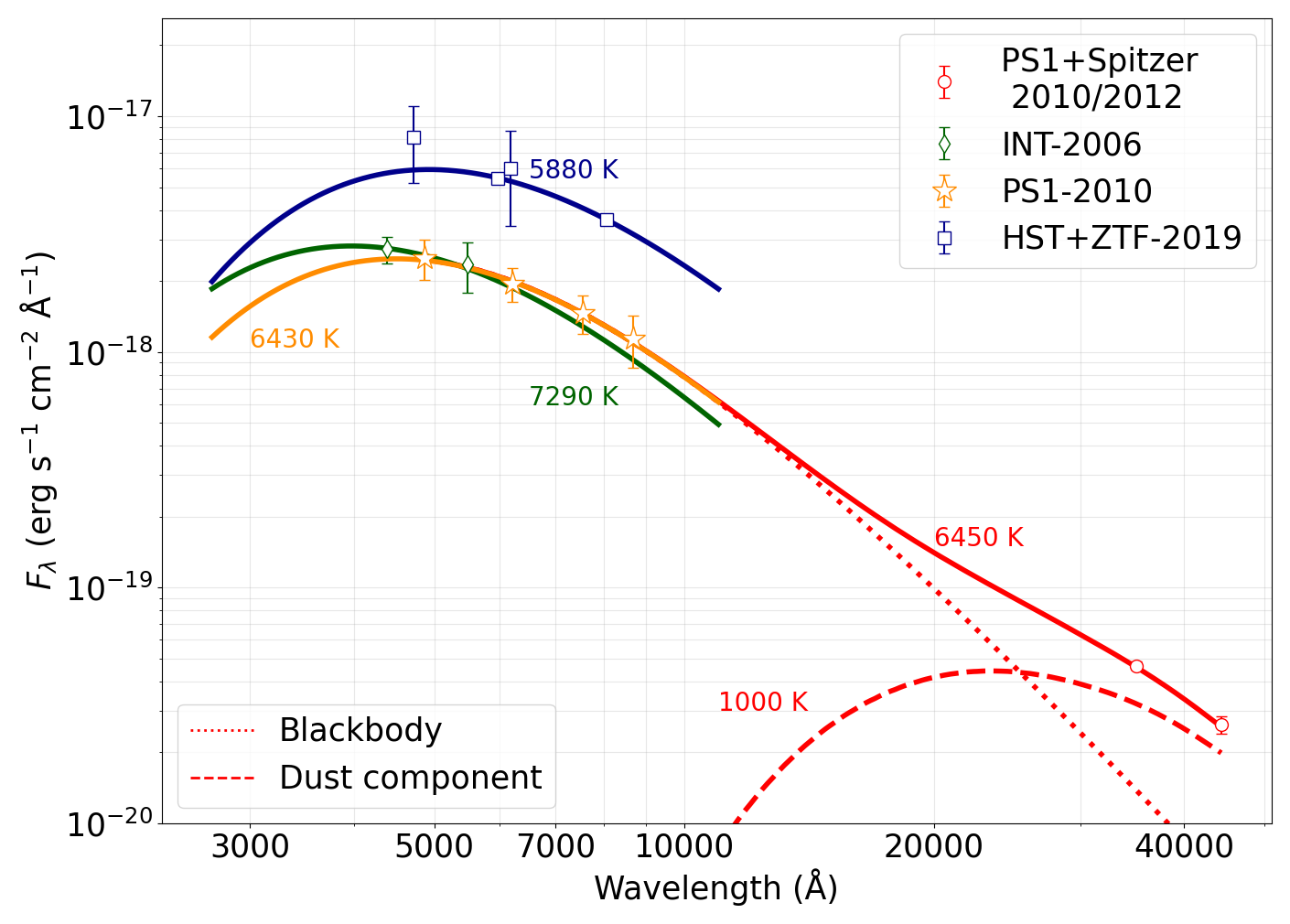}
    \caption{Blackbody fits to the multi-band observations of AT2021blu, illustrating the SED evolution across the 2006, 2010, 2010–2012, and 2019 epochs (top panel), and using ZTF forced photometry shown in Fig. \ref{fig:forced} for the 2018, 2019, and 2020 epochs (bottom panel).}
    \label{fig:SED_evol}
\end{figure}

In the case of AT2021blu, \cite{2023A&A...671A.158P} identified a source with modest variability in ground-based imaging from 2006–2016, and used it to perform a blackbody fit. We performed a blackbody fit on the same data and found very similar values\footnote{The authors found $T_{eff}=6800\pm300$K, $L_{bol} = (4.1 \pm 0.6) \times 10^{4}$ $L_{\odot}$ and $R = 144 \pm 14 R_{\odot}$}: $T_{eff}=6751_{-601}^{+780}$K, $L_{bol} = (4.1 \pm 0.3) \times 10^{4}$ $L_{\odot}$ and $R=148_{-29}^{32} R_{\odot}$. These estimates correspond to an epoch $5-15$ years before the outburst. In order to study the evolution of the blackbody temperature across the different epochs, we also performed independent blackbody fits to the photometric observations obtained in 2006, 2010, and 2019. For the 2006 epoch, we used the PS1 g, r, i, and z magnitudes, while for the 2010 epoch we adopted the B- and V-band magnitudes from INT/WFC as reported by \citet{2023A&A...671A.158P}. Following \citet{2020MNRAS.496.5503B}, we also fitted the PS1 2010 and Spitzer 2012 photometry assuming a blackbody for the central source and an optically thin dust component \cite[see Eq. 1--3 in][]{2020MNRAS.496.5503B}. The dust temperature, central source radius, and dust mass were constrained via MCMC to reproduce the infrared-extended SED. For the 2019 epoch, the fit was performed using the HST detection together with interpolated ZTF forced-photometry at the same epoch, which provide g- and r-band magnitudes of $22.702\pm0.198$ and $21.731\pm0.308$ mag, respectively.
The results indicate cooling during the precursor phase, as it can be seen in Fig. \ref{fig:SED_evol}. The blackbody and dust parameters are listed in Table \ref{tab: BBfit}. 
\begin{table*}
\centering
\renewcommand{\arraystretch}{1.3}
\begin{tabular}{lccccc}
\hline
\hline
Source & $T_{\rm eff}$ (K) & $L$ ($\times 10^{5}L_\odot$) & $R_{*}$ ($R_\odot$) & $M_{\rm dust}$ ($\times 10^{-7}M_\odot$) & $T_{\rm dust}$ (K) \\
\hline
HST+ZTF\tablefootmark{a}
-2019 &
$5880^{+309}_{-260}$ &
$1.04^{+0.03}_{-0.03}$ &
$310.3^{+27.8}_{-27.7}$ &
$\cdots$ &
$\cdots$ \\
PS1+SPITZER\tablefootmark{a}
-2010/2012&
$6450^{+1264}_{-984}$ &
$0.5^{+0.7}_{-0.3}$ &
$160^{+61}_{-33}$ &
$1.02^{+0.84}_{-0.38}$ &
$1000^{+135.21}_{-141.63}$\\
PS1-2010 &
$6430^{+1182}_{-867}$ &
$0.40^{+0.52}_{-0.44}$ &
$160.3^{+51.4}_{-40.9}$ &
$\cdots$ &
$\cdots$ \\
INT-2006 &
$7290^{+4521}_{-1839}$ &
$0.45^{+0.18}_{-0.07}$ &
$125.0^{+133.0}_{-72.5}$ &
$\cdots$ &
$\cdots$ \\
\hline
\end{tabular}
\caption{Blackbody fit parameters for AT2021blu at the different detection epochs: ground-based INT (2006) and Pan-STARRS1 (2010), as well as the HST+ZTF (2019) and PS1+Spitzer (2010/2012) datasets. We also list the dust-component parameters adopted to fit the PS1+Spitzer SED.
\tablefoottext{a}{These data are new and unpublished.}
}
\label{tab: BBfit}
\end{table*}

We finally adopted the blackbody fit estimates obtained from the \cite{2023A&A...671A.158P} photometry to place AT2021blu on the HR diagram, as they are based on photometry aggregated over a long baseline (2006–2016) and are consequently more robust than the forced-photometry measurements or the HST detection alone. Moreover, during this interval the source was likely in a quiescent (precursor) state, whereas by the time of the 2019 HST observation, it was already approaching the onset of the outburst. 

\section{Progenitor modeling} 
\label{sec:modeling}

As discussed in Sect.~\ref{sec:data description}, the archival data were obtained less than two decades before the outburst and therefore likely capture the system during a phase of unstable mass transfer. Our approach therefore focused on identifying progenitor configurations that both match the observations and are undergoing such an unstable mass-transfer regime.

\subsection{Stellar models}

To constrain the binary progenitor properties of AT2021biy and AT2021blu—in terms of donor mass ($M_\mathrm{d}$), mass ratio ($q=M_{d}/M_{a}$ with $M_{a}$ the companion mass), and initial orbital period ($P$)—we extended the approach of \cite{Blagorodnova2021} by incorporating mass-transfer instability criteria. We first computed a grid of binary evolution models spanning a range of ($M_\mathrm{d}, q, P$) combinations (detailed in Sect. \ref{sec:results}) and identified those whose evolutionary tracks cross the observed locations of each event in the HR diagram during the mass-transfer phase. From this subset, we then selected the systems that also satisfy the mass-transfer instability criteria, thereby isolating those likely to be close to a runaway episode leading to a CE phase.

We used version 11554\footnote{\url{https://zenodo.org/records/2630923}} of the stellar-evolution code MESA \citep{2011ApJS..192....3P,2013ApJS..208....4P,2015ApJS..220...15P,2018ApJS..234...34P,2019ApJS..243...10P}, with the \texttt{mesasdk-x86\_64-linux-20190404} SDK, to link the observed luminosity and temperature derived in Sect. \ref{sec: archival_data} to binary progenitors. This code follows the key parameters that govern binary interactions and system evolution. 

\subsubsection{Physical ingredients}
\label{sec:physical_in}

The physical ingredients used in our modeling are inspired by previous works \citep{2020A&A...638A..55K,Blagorodnova2021,2022A&A...661A.123F}. Convection is modelled with the classical \cite{1947ApJ...105..305L} criterion and the standard mixing-length theory introduced by \citet{1958ZA.....46..108B}, with the length parameter $\alpha=1.5$. We adopted a highly efficient semiconvective mixing parameter, $\alpha_{sc}=100$ and allowed for core overshooting in both hydrogen- and helium-burning cores ($f_{ov}=0.345$) while neglecting overshooting in shell-burning regions and envelope convection zones. Such assumptions are consistent with the mixing requirements inferred for low-metallicity massive stars \citep[see, e.g.,][]{2019A&A...625A.132S}. Stellar winds follow \citet{2011A&A...530A.115B} using prescriptions from \citet{1990A&A...231..134N,1995A&A...299..151H,2000A&A...362..295V,2001A&A...369..574V} depending on hydrogen abundance and effective temperature. The nuclear reactions are assumed to follow the standard MESA networks (\texttt{basic.net} for H and He burning and \texttt{co\_burn.net} for C and O burning) and we used blended equation-of-state (EOS) and opacity tables based on \citet{1995ApJS...99..713S,1996ApJ...464..943I,2005ApJ...623..585F,2010CoPP...50...82P}. A detailed description of the physical ingredients is provided in Appendix \ref{app:phys_ingredients}.

For each binary system, the model starts from a previously constructed zero-age main-sequence (ZAMS) massive donor at the metallicity of each transient. We then added a point-mass companion and assumed it does not accrete any mass. Despite observational evidence that at least in some cases mass transfer may be predominantly conservative \citep[e.g.][]{2025ApJ...990L..51L}, we have chosen for the sake of simplicity to assume fully non-conservative mass transfer. Only the evolution of the primary star is modeled here, while the secondary star merely channels the mass transfer. No initial rotation is imposed, but gravitational interaction and tidal forces induce rotation during the modeling. We allowed differential internal rotation in the modelling of the donor star, including various angular-momentum transport processes within the stellar interior (see Appendix \ref{app:phys_ingredients}). 

Following observational evidence of disks and jets present in actively mass transferring binaries such as $\beta$ Lyrae  \citep{1996A&A...312..879H,2003A&A...405..223L}, we assumed that half of the transferred mass is lost from the vicinity of the secondary star through isotropic re-emission (IRE) while the remaining half is ejected into a coplanar circumbinary toroid (CT) with radius $a_{r}=\gamma^{2}a$, where $\gamma=1.2$ and $a$ denotes the orbital separation. The resulting effective specific angular-momentum loss $\eta_{\rm eff}$ can be expressed in units of the system’s orbital specific angular momentum $J/M$ with $M=M_{d}+M_{a}$ the total mass of the binary system \citep[see][]{1997A&A...327..620S}:

\begin{equation}
\label{eq:alpha_eff}
\eta_{\rm eff}
= \frac{\eta_{\rm CT}+\eta_{\rm IRE}}{2}
= \frac12\left(\frac{(1+q)^2}{q}\gamma+q\right)
\end{equation}

We adopted the \cite{1990A&A...236..385K} mass-transfer scheme in the optically thick Roche Lobe overflow case. The termination condition in our models is set when ${}^{4}$He is almost totally depleted in the donor star core, at which point most of the mass transfer is expected to have been completed.

All the MESA input files used in this work, along with the corresponding outputs, are available online\footnote{DOI: \url{10.5281/zenodo.18167185}}.

\subsection{Mass transfer instability}
\label{sec:theory}

When the donor star enters the Roche-lobe overflow (RLOF) phase, the fate of the system strongly depends on the ensuing mass-transfer rate, in particular on whether it remains stable or runs away. For this reason, incorporating mass-transfer instability is crucial when searching for progenitor configurations that can lead to a CE phase. As we mentioned in Sect. \ref{sec: archival_data}, the observational data used for placing the progenitor systems on an HR diagram, date from less than two decades before the detected outburst, most likely corresponding to a phase of already unstable mass transfer. We therefore applied the instability criteria presented in this section to constrain the progenitor properties.

The stability of mass transfer primarily depends on how the donor and the companion respond to the mass-loss rate, particularly through the evolution of their radii and their respective Roche lobes. The classical criterion \citep[e.g.][]{1985ibs..book...39W,1997A&A...327..620S} for describing instability in a binary system involves comparing the adiabatic response of the donor star's radius to mass loss $\zeta_{ad}=d\log R_{d}/d \log M_{d}$ with that of its Roche lobe $\zeta_{L_{1}}=d\log R_{L_{1}}/d \log M_{d}$. If $\zeta_{ad}<\zeta_{L_{1}}$, the donor star cannot remain confined within its Roche lobe, leading to an unstable mass transfer on a dynamical timescale and the formation of a CE. Nevertheless, as pointed out by \citet{2002ApJ...565.1107P, 2011ApJ...739L..48W, 2012ApJ...744...52P, 2015MNRAS.449.4415P, 2023A&A...669A..45T}, the ability of the outer layers to thermally readjust, rather than responding purely adiabatically on dynamical timescales, can help stabilise the mass transfer. Furthermore, donors with radiative envelopes typically transfer a substantial amount of mass before entering a runaway phase (the so-called delayed dynamical instability, see also Sect. \ref{sec:results}). This is qualitatively consistent with the large amount of slow moving gas detected in the vicinity of AT2021biy and AT2021blu (see Paper II), and in general from the LRNe literature \citep[e.g.,][]{2022A&A...664A..12M}.

We describe mass-transfer stability in our models according to the three criteria presented in \cite{2023A&A...669A..45T}. First, we use the quasi-adiabatic (QA) criterion, measuring the donor's ability to thermally readjust its outer layers in response to mass loss. Specifically, when the mass loss rate exceeds the critical rate defined by the local thermal adjustment timescale of the outer envelope, $\dot{M}_{d}>\dot{M}_{\mathrm{th,crit}}$, the donor transitions to a fully adiabatic response. The critical thermal mass-loss rate is expressed as:
\begin{equation}
\left\{
\begin{aligned}
    \dot{M}_{\mathrm{th,crit}} &= \max_m\left(\frac{m - M}{\tau_{\mathrm{th}}(m)}\right) \,\\
    \tau_{\mathrm{th}}(m) &= \frac{1}{L} \int_{m}^{M} c_p(\tilde{m})\, T(\tilde{m}) \, d\tilde{m} \,
\end{aligned}
\right.
\end{equation}
where $m$ is the mass coordinate, $M$ is the mass at the donor's surface, $c_p$ is the specific heat at constant pressure, $T$ the temperature, $L$ the surface luminosity, and $\tau_{\mathrm{th}}(m)$ the local thermal timescale of a layer at mass coordinate $m$.

Second, we use the outer lobe overflow (OLOF) criterion since angular momentum loss through either of the Lagrangian points $L_{2}$ or $L_{3}$ during mass transfer can significantly shrink the binary orbit \citep[e.g.][]{2017ApJ...850...59P,2020ApJ...893..106M} and lead to a dynamical instability. However, as noted in \cite{2010A&A...521A..81B}, this condition does not necessarily guarantee the onset of a runaway situation. Nevertheless, during OLOF the donor star deviates substantially from spherical symmetry, making its representation in one dimension increasingly challenging.
The onset of OLOF corresponds to the condition \(R_{\mathrm{d}} > R_{\mathrm{L_3}}\), where $R_{L_{3}}$ is the equivalent-volume radius of the outer lobe. We use the fitting formula provided by \citet{2023A&A...669A..45T} to compute $R_{L_{3}}$ relative to the equivalent-volume radius  of the Roche lobe $R_{L_{1}}$ \citep{1983ApJ...268..368E}:
\begin{equation}
\label{eq:RL}
\left\{
\begin{aligned}
     && R_{L_3} \approx R_{L_{1}}\left(1 + \frac{0.441 q^{0.325}}{1 + 0.412 q^{0.8}}\right)\\
     && \frac{R_{\mathrm{L_1}}}{a} \approx \frac{0.49q^{2/3}}{0.6q^{2/3} + \ln(1+q^{1/3})}
\end{aligned}
\right.
\end{equation} 

Finally, the third criterion used to assess mass transfer instability measures when mass transfer accelerates to a dynamical timescale, i.e. whether the relative change in either the orbital distance or the donor mass over one orbital period exceeded the threshold $A_{dyn}$: 
\begin{equation}
    \max\left( \left| \frac{\dot{M}_{\mathrm{d}}}{M_{\mathrm{d}}} \right|, \left| \frac{\dot{a}}{a} \right| \right) \cdot P > A_{\mathrm{dyn}},
 \label{eq: dyna_crit} \\
\end{equation}
where $a$ is the orbital separation, $\dot{a}$ and $\dot{M}_{d}$ correspond to variations over one orbital period $P$.
We assume $A_{dyn}=0.05$, obtained from simulations and corresponding to the onset of the plunge-in phase \citep[see][]{2020cee..book.....I}. This criterion corresponds to a phase where the Roche model—and with it, any one-dimensional representation—ceases to be valid. This stage represents the ultimate point of instability, beyond which the onset of a CE phase becomes unavoidable.  

In addition to these mass-transfer instability criteria, we also checked whether the Darwin instability occurred. This instability arises when the total spin angular momentum of the stellar components exceeds a critical fraction of the orbital angular momentum. Beyond this threshold, the orbit can no longer supply enough angular momentum to maintain synchronism, and tidal coupling drives a runaway inspiral, leading to rapid orbital shrinkage and possibly a CE phase \citep[see][]{1879Obs.....3...79D}. Since this criterion becomes relevant only for high mass ratios ($q\geq10$), we do not expect it to operate over the full range of our progenitors.

We considered that mass transfer instability occurs in the binary as soon one of these conditions is satisfied. The QA criterion best describes the actual transition to runaway mass transfer found in detailed binary evolution calculations by \cite{2023A&A...669A..45T}, while the OLOF and dynamical criteria provide additional, complementary perspectives. However, as previously mentioned, relying on MESA predictions becomes risky once the assumptions underpinning a one-dimensional representation, in particular spherical symmetry, breaks down. We were therefore cautious in interpreting any criterion that occurs after OLOF.

\subsection{First selection based on observations}
\label{sec:results}

To find a suitable match for our progenitor systems, we explored a wide range of donor star masses—from 11$M_{\odot}$ to 30$M_{\odot}$ for AT2021blu, and from 17$M_{\odot}$ to 40$M_{\odot}$ for AT2021biy. The choice to look for higher masses for AT2021biy was motivated by both photometric and spectroscopic observations. The measured ejecta velocity ranges and luminosities suggest a more energetic event and thus a likely more massive system for AT2021biy than for AT2021blu \citep[see][]{Cai2022AA,2023A&A...671A.158P}. In order to find a progenitor whose evolution in the HR diagram matches the observational constraints during the mass-transfer phase, we covered mass ratios up to 40 and initial binary separations between 200 $R_{\odot}$ and 2000 $R_{\odot}$ for both objects. We describe in this section our first selection of candidates from the full set of modelled progenitor systems. All identified progenitors fill their Roche lobe only after core hydrogen exhaustion, as they cross the Hertzsprung gap. This classifies them as case B mass transfer systems, characterized by envelopes that remain predominantly radiative during the mass transfer phase. 

Within our grid of MESA models, we found several candidates whose evolution track passes through the observed region in the HR diagram for AT2021biy and AT2021blu during mass transfer. Based on these grids, we can already place strong constraints on the set of progenitors by excluding cases where the donor-star mass is either too low—so the luminosity of the progenitor is always less than the observed luminosity (setting lower limits on the progenitor masses of $M_{d} > 18M_{\odot}$ for AT2021biy and $M_{d} > 13M_{\odot}$ for AT2021blu)—or too high and it cannot reach that region before entering a CE phase. Furthermore, as the progenitor mass increases, reproducing the observed $(T_{\rm eff}, L)$ coordinates requires a larger initial binary separation, which consequently shifts the onset of RLOF to lower effective temperatures. This sets upper limits on the progenitor masses of $M_{d}\lesssim26 M_{\odot}$ for AT2021biy and $M_{d}\lesssim15 M_{\odot}$ for AT2021blu (see Sect. \ref{sec:flattening}). In Figs. \ref{fig:AT2021biy_grid} and \ref{fig:AT2021blu_grid} in the Appendix we present a subset of the calculated grid, showing the region of parameter space in which we find matching models.

\subsection{Impact of initial binary separation on the HR diagram evolution during mass transfer}
\label{sec:flattening}

\begin{figure*}[h!]
    \centering
    \includegraphics[width=\textwidth]{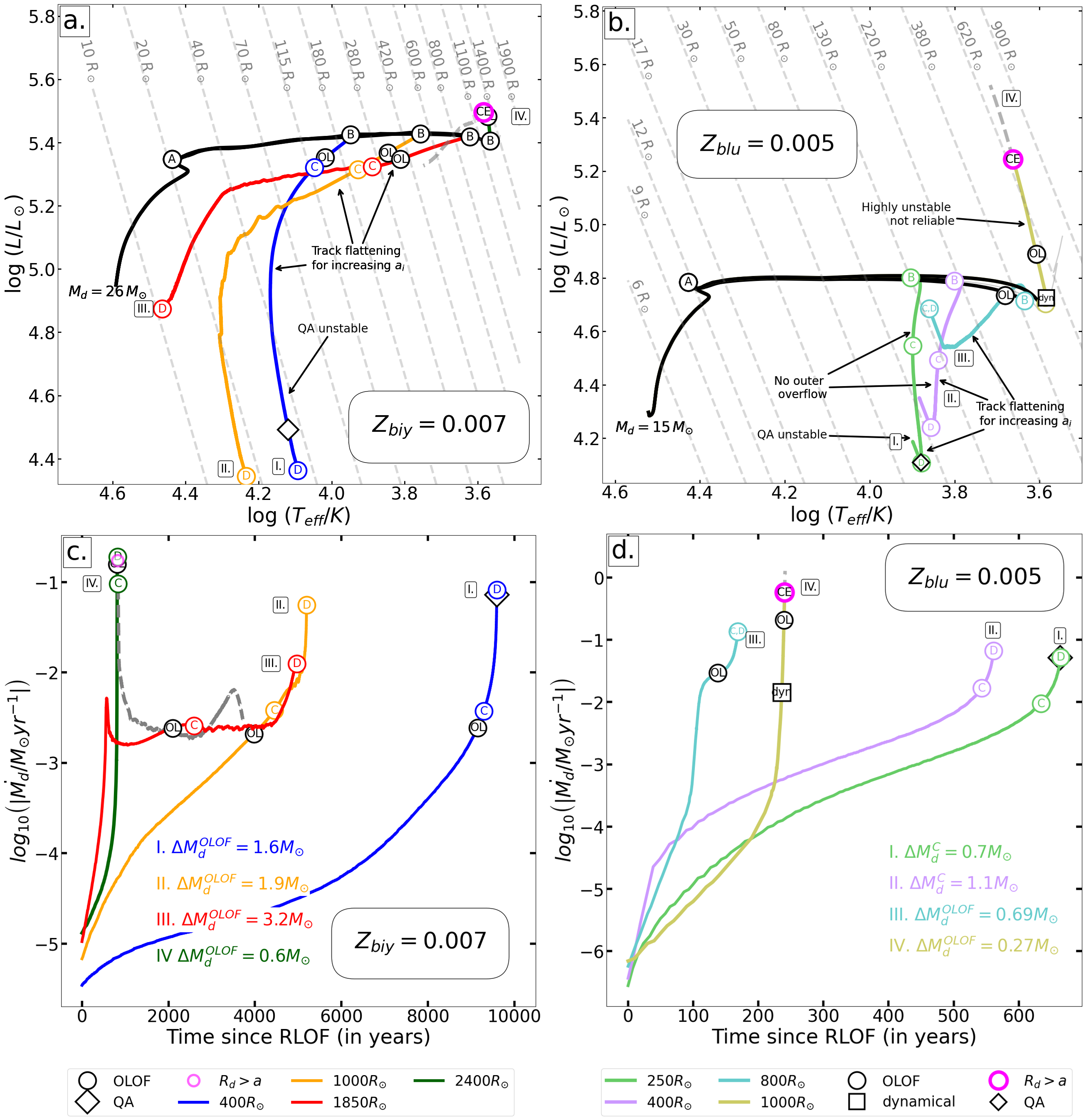}
    \caption{\textit{Top panels}: The main types of stellar evolution for the AT2021biy and AT2021blu progenitor systems found in Figs. \ref{fig:AT2021biy_grid} and \ref{fig:AT2021blu_grid}. We display the following key moments in the evolution: \textbf{A}) the end of MS phase, \textbf{B}) the onset of RLOF, \textbf{C}) the maximum overfilling factor and \textbf{D}) the moment of $\max \dot{M}_{d}$. Single-star models of the progenitors are indicated with gray lines. \textit{Bottom panels}: The associated mass lost rates for each transient. We also display the total mass loss from the onset of RLOF to the onset of OLOF ($\Delta M_{d}^{OLOF}$) or to the maximum overfilling factor ($\Delta M_{d}^{C}$) for each type of stellar evolution . All these models correspond to a mass ratio $q=5$, with $M_{d}=26M_{\odot}$ for AT2021biy and $M_{d}=15M_{\odot}$ for AT2021blu. The time axis of the two bottom panels starts when the mass-transfer rate exceeds the wind mass-loss rate for the first time.}
    \label{fig:SE_types}
\end{figure*}

In our grid of models, we found that, for the higher donor-mass range, increasing the initial orbital separation produces a flattening of the stellar evolutionary tracks at early stages during mass transfer. In Fig. \ref{fig:SE_types}, we illustrate this behaviour and relate it to the evolution of the mass-loss rate for each binary configuration. For both transients, variations in effective temperature are accompanied by only modest changes in luminosity at larger initial binary separations (see Figs. \ref{fig:SE_types}a and \ref{fig:SE_types}b). We defined the onset of mass transfer as the first time the mass-transfer rate exceeds the wind mass-loss rate $\dot{M}_{d}>\dot{M}_{\rm wind}$ \citep[see][]{2023A&A...669A..45T}.  

For small initial orbital separations (see track I in Figs. \ref{fig:SE_types}a and \ref{fig:SE_types}b), the donor undergoes RLOF while it still has a radiative envelope and thereby contracts under mass loss. As the envelope is stripped at an increasingly high rate, the energy required to lift the exposed deeper layers to the surface lowers the luminosity, resulting in an almost vertical evolution in the HR diagram. 

As the initial binary separation increases (see tracks I–III in Figs. \ref{fig:SE_types}a and \ref{fig:SE_types}b), the evolutionary tracks during the early phases of mass transfer become progressively flatter, although at later times they show a sharp decline in luminosity. This behaviour is especially evident in Fig. \ref{fig:SE_types}a for AT2021biy. The time evolution of the mass-transfer rates shows a corresponding dependence on the initial binary separation (see Figs. \ref{fig:SE_types}c and \ref{fig:SE_types}d), with the rise in the mass-transfer rate becoming progressively less gradual at larger separations. This can be explained by the fact that wider systems reach the onset of RLOF at lower effective temperatures (e.g. $\sim$ 4000 K for track III in Fig. \ref{fig:SE_types}a), allowing a shallow outer convective layer to form. This produces an increasingly strong response of the donor to mass loss during the early stages of mass transfer.

At even larger initial separations (track IV in Figs. \ref{fig:SE_types}a and \ref{fig:SE_types}b), the donor exhibits a much more dramatic response to mass loss. In this case, only a few hundred years after the onset of RLOF, the mass transfer transitions to a dynamical timescale, and the donor radius rapidly exceeds the binary separation ($R_{d}>a$). The system is therefore expected to enter a CE phase almost immediately, without passing through a prolonged phase of quasi-stable mass transfer. Given the limitations of MESA in this regime, we excluded models whose evolutionary tracks intersect the observed HR region only after the onset of dynamical instability. 

Importantly, this delayed runaway of $\dot{M}_{d}$ (see Figs. \ref{fig:SE_types}c and \ref{fig:SE_types}d) together with the associated flattening of the evolutionary tracks (see \ref{fig:SE_types}a and \ref{fig:SE_types}b) explains why we did not find solutions matching the observed HR location for donor masses $M_{d}\gtrsim26 M_{\odot}$ for AT2021biy and $M_{d}\gtrsim15 M_{\odot}$ for AT2021blu, even when increasing the initial binary separation.

As a final point, the mass-transfer-rate evolution of track III for AT2021biy (Fig. \ref{fig:SE_types}c) deserves particular attention. It appears to represent a transition between the gradual rise seen in tracks I and II and the early runaway behaviour of track IV. In this configuration, only a few hundred years after the onset of RLOF, the system reacts strongly to mass loss, producing an initial peak in the mass-transfer rate, likely due to the presence of an outer convective layer. As this layer is rapidly stripped, the newly exposed radiative layers cause a sharp readjustment of the mass-transfer rate, which remains below $\sim10^{-2}M_{\odot} yr^{-1}$. This temporarily stabilized phase is followed, a few thousand years later, by a renewed increase in mass-transfer rate resembling the late-time evolution of the other systems (see tracks I, II, and IV). This suggests that, regardless of the initial separation, we expect all these systems to ultimately evolve toward an unstable regime.

\subsection{Second selection imposing instability criteria}
\label{sec:instability}

\begin{figure*}[h!]
    \centering
    \includegraphics[width=\textwidth]{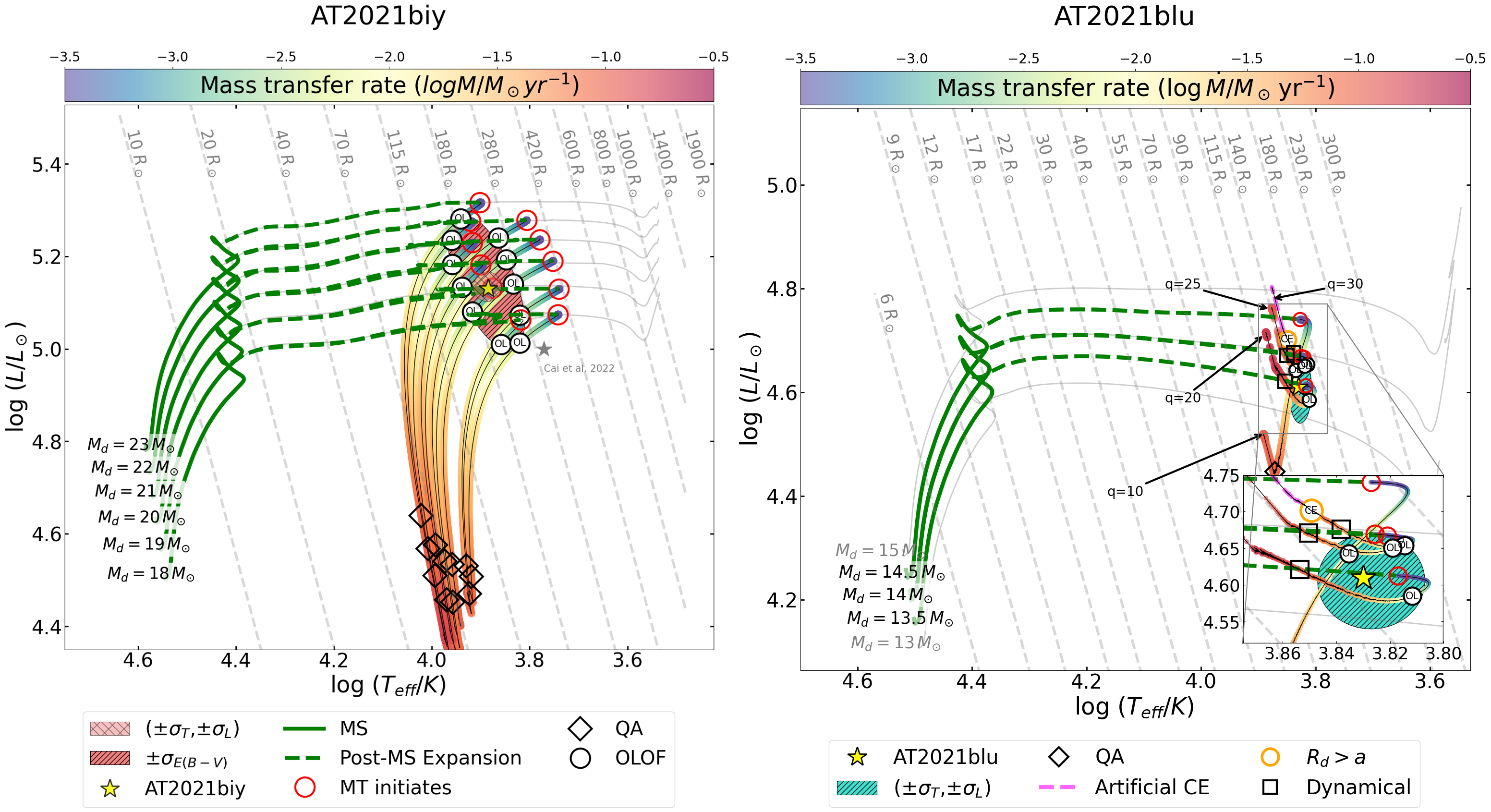}
    \caption{Progenitor systems that match the observed HR location while undergoing OLOF are shown here for the two transients, AT2021biy and AT2021blu. The mass transfer phase begins at the red circle, while the OLOF occurs at the black circle. Single-star models of the progenitors are indicated with gray thin lines. In the case of AT2021biy, all the displayed progenitors correspond to a mass ratio $q=10$. In the case of AT2021blu, beyond the point where $R_{d}>a$ (indicated by the CE symbol), the stellar evolution is no longer physically meaningful, as it would correspond to an artificial CE phase that is not self-consistently modelled by MESA.}
    \label{fig:candidates}
\end{figure*}

As discussed above, mass-transfer instability lies at the heart of CE formation and must therefore be incorporated into the progenitor modelling.

In all our models, the Darwin instability criterion is only met at very large mass ratios, $q \gtrsim 25$, for AT2021biy, and it is never satisfied for AT2021blu over the range of mass ratios explored (extending up to $q=40$). Note that in our models, this criterion is typically fulfilled only in configurations in which the donor expands to $R_{d}>a$ and therefore only once the system is already dynamically unstable. We tested models in which the donor star is initialized at 50\% of its critical rotation rate $\Omega_{\text{crit}}$ at the ZAMS. This has little impact on most instability criteria, but leads to an earlier onset of the Darwin instability.

The OLOF criterion is typically met significantly earlier than the QA threshold (e.g., see Fig. \ref{fig:SE_types}), in contrast to the findings of \citet{2023A&A...669A..45T} for lower-mass binaries (2–8 $M_{\odot}$) and intermediate mass ratios ($q\leq10$). This is partly because, for more massive donors, the evolution timescales are much shorter and the corresponding mass-transfer rates are higher, so that larger overfilling factors are required to reach the necessary $\dot{M}_{th, crit}$. 
Furthermore, we found that the higher the mass ratio, the sooner the OLOF criterion is satisfied after mass transfer begins. This can be attributed to two factors: (i) the dependence of the ratio $R_{L_{3}}/R_{L_{1}}$ on $q$ (see Eq. \ref{eq:RL}), which decreases towards unity as $q$ increases, such that less radial expansion beyond the $L_{1}$ point is required to reach the $L_{3}$ surface and trigger OLOF; (ii) higher-$q$ systems typically undergo more rapid, unstable mass transfer due to a stronger orbital shrinkage, which causes $R_{L_{1}}$ to shrink and the degree of overfill to grow on short timescales.
In our massive binary systems with $q\gtrsim10$, as soon as mass transfer begins, even a very small overfilling factor is sufficient for the stellar radius to reach the outer Lobe, so OLOF is triggered at early times while the donor still responds non-adiabatically and the QA criterion is not yet satisfied. 

Conversely, as the mass ratio approaches unity, the mass transfer becomes increasingly stable. This is because the orbit does not contract as severely in response to mass transfer, which prevents the runaway feedback loop between mass loss and Roche-lobe shrinkage. Consequently, even though systems with $q\lesssim2$ do undergo OLOF, the subsequent interaction remains relatively mild and the mass-transfer rate stabilizes (see case III in Fig. \ref{fig:SE_types}a). This behaviour was already noted in Sect. \ref{sec:theory}, where we emphasized that the onset of OLOF does not necessarily imply an immediate transition to a dynamical phase. As shown in Sect. \ref{sec:LC_models}, these configurations with $q\lesssim2$ are also inconsistent with the observed light curves, and we therefore discard them as viable progenitors. This caveat is further discussed in Sect. \ref{sec:discussion}.

For each instability criterion, we searched for systems that satisfy it while they are crossing the observed HR region. In practice, the viable solutions for both AT2021blu and AT2021biy are predominantly found near the onset of OLOF. We therefore adopted this criterion as the primary guide for our progenitor selection. 

For AT2021biy, we did not identify any progenitor models that satisfy the QA criterion while remaining consistent with the HST pre-outburst constraints. Reproducing the HST detection under the QA condition would require donor masses $\geq30M_{\odot}$. In this mass range, increasing the initial binary separation leads to a more horizontal evolution in the HR diagram, as discussed in Sect. \ref{sec:flattening}, such that no viable progenitor solutions are found. In contrast, adopting the OLOF criterion yields solutions consistent with the observations for $M_{d}=18-23M_{\odot}$, $q\gtrsim4$ and $a=300-700R_{\odot}$. We analyzed the properties of these systems and illustrate the specific case of $q=10$ in Fig. \ref{fig:candidates} (left panel). For this mass ratio, we show, for each donor mass, the models with the minimum and maximum initial separations that place the onset of OLOF within the 1$\sigma$ observed HR region. 

\begin{figure}[h!]
    \centering
    \includegraphics[width=\linewidth]{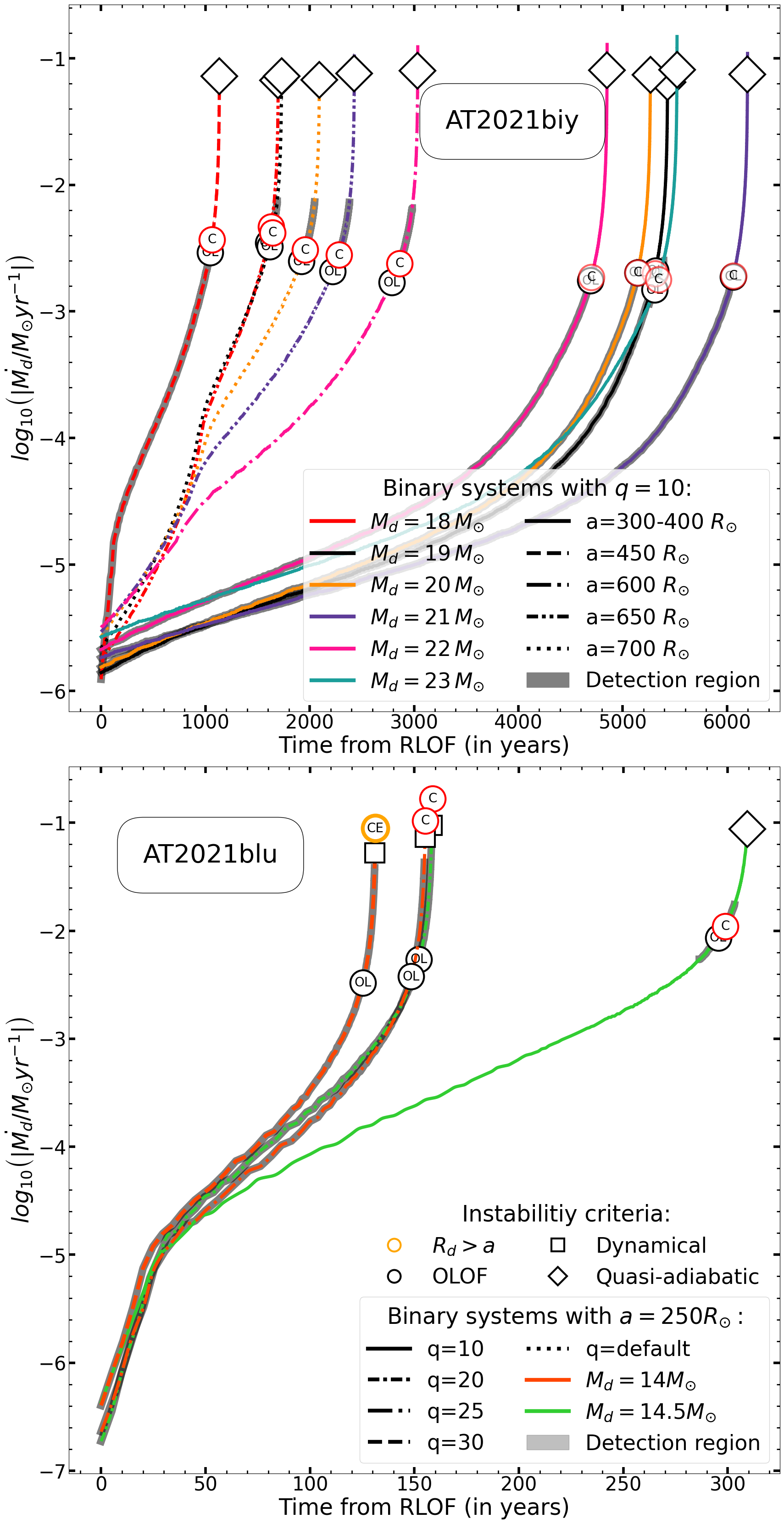}
    \caption{Mass-transfer rates of the progenitor configurations shown in Fig.~\ref{fig:candidates} for the two transients AT2021biy and AT2021blu, together with the locations of the instability criteria. We display by \textbf{C}) the moment of the maximum overfilling factor. The shaded gray lines highlights the observed HR location shown in Fig. \ref{fig:candidates}. The time axis of the two panels starts when the mass-transfer rate exceeds the wind mass-loss rate for the first time.}
    \label{fig:candidates_Mdot}
\end{figure}

Fig. \ref{fig:candidates_Mdot} presents the evolution of the mass-transfer rates for the models shown in Fig. \ref{fig:candidates}. As illustrated in Fig. \ref{fig:candidates_Mdot}, the mass-transfer rates at the epoch when the models simultaneously reach the HST detection region and the onset of OLOF are in the range $10^{-3}-10^{-2} M_{\odot} yr^{-1}$, corresponding to a total transferred mass of $0.4 - 0.6M_{\odot}$ since the onset of the RLOF. The corresponding amounts of mass loss by stellar wind as well as during RLOF are presented in Fig. \ref{fig:candidates_ML} in the Appendix. Owing to our fully non-conservative assumption, the computed amounts of transferred mass leaving the system should be interpreted as an upper limit. 

In the case of AT2021blu, we found progenitors that agree with the observed HR region at the onset of OLOF while they are getting close to instability according to the QA criterion and even the dynamical criterion (see the right panel of Fig. \ref{fig:candidates}). Because the observed region in the HR diagram is less extended than for AT2021biy, we derived tighter constraints on the progenitor systems, with donor masses of $M_{d}=14 \pm 0.5M_{\odot}$, mass ratios $q\gtrsim10$, and orbital separations in the range $a=200-300R_{\odot}$. The mass-transfer rates at the epoch when the progenitor evolution matches the observational constraints in the HR diagram and at the onset of OLOF, are $10^{-3}-10^{-2} M_{\odot} yr^{-1}$ (see Fig. \ref{fig:candidates_Mdot}), corresponding to a total expelled mass of $0.03 - 0.28M_{\odot}$ for AT2021blu since the onset of the RLOF (see Fig. \ref{fig:candidates_ML}).

\section{Ejecta mass estimates: energy budget and light-curve modeling}
\label{sec:energetics}

Now that we have modelled the progenitor system during the pre-outburst phase with MESA, we can connect it to the outburst through the dynamical phase. We adopted an energetic approach to estimate the envelope mass that would be ejected during a CE event, assuming the interaction begins at the time of OLOF. In this section, we compare the energy required for such an ejection with the energy provided by the system’s dynamics, in order to estimate the corresponding ejected envelope mass. 

We computed both lower and upper limits based on two different assumptions regarding the contribution of orbital energy after the point at which energy sources and sinks balance. The lower limit is obtained by adopting a local energy-deposition scheme, in which the orbital energy released during inspiral is deposited into, and immediately used to unbind, the envelope layers exterior to the companion’s instantaneous orbit. This approach is based on that of \cite{Blagorodnova2021} and assumes that, once the instantaneously released energy becomes insufficient to unbind subsequent layers, any orbital energy released at smaller radii does not contribute to further envelope ejection. By neglecting these late-stage energy contributions, this approximation provides a lower limit.

To complement this estimate, we also derived an upper limit on the total ejecta mass, for which we assumed that all orbital energy released up to the point where the donor’s core fills its Roche lobe contributes to unbinding the envelope. In other words, we solved the energy conservation equation by including the full available energy up to the point at which we expect a merger between the companion and the donor’s helium core.

The derivation of the lower limit implicitly assumes that the released orbital energy is instantaneously and locally used to unbind the envelope layers. In practice, however, a finite time delay exists between energy injection and the physical ejection of the envelope, implying that the energy is redistributed throughout the envelope rather than deposited locally.
Our upper-limit estimate partially accounts for this non-local energy redistribution, but in a simplified manner. We address these limitations in Sect.~\ref{sec:discussion}, where we propose a more sophisticated approach that explicitly incorporates these effects into the modeling. To keep things simple, we modelled the optimal case with no energy losses, implicitly assuming a perfectly efficient conversion ($\alpha_{CE}=1$). 

Furthermore, our massive Hertzsprung-gap progenitors ($M_{d}\geq 10 M_{\odot}$) have tightly bound, low-density envelopes \citep{2003MNRAS.341..385P} and we expect the energy generated by accretion onto the companion to be negligible. In addition, for high mass ratios ($q\gtrsim10$) the inspiral should be rapid, so the time-integrated accretion contribution should also be negligible. Accordingly, we did not include any accretion-powered contribution in our energy budget.

We adopted the following standard form for the energy conservation equation \citep[see, e.g.][]{1984ApJ...277..355W,1988ApJ...329..764L,2013A&ARv..21...59I, 2020cee..book.....I}: 
\begin{equation}
\label{eq:energy_cons}
\Delta E_{\mathrm{orb}} = E_{\mathrm{ej}} + E_{\mathrm{bind}}
\end{equation}
where each term can be expressed as a function of the lost envelope mass:
\begin{subequations}
\label{eq:energy_terms}
\begin{align}
\Delta E_{\text{orb}}(M_{env}) &= -\frac{G M_{d,ce} M_{a}}{2 a_{OLOF}} 
+ \frac{G (M_{d,ce} - M_{\text{env}}) M_{a}}{2 r(M_{env})}
\label{eq:E_orb}\\
E_{\text{bind}}(M_{env}) &= - \int_{M_{d,ce} - M_{\text{env}}}^{M_{d,ce}} 
\left( -\frac{Gm}{r(m)} + u(m)\right) dm \\
E_{ej}(M_{env}) &= \frac{1}{2} v_{\text{ej},\infty}^2 M_{\text{env}} 
\label{eq:bulk_kinetic}
\end{align}
\end{subequations}

We assumed that the CE phase initiates at the onset of OLOF in our MESA models, so $M_{d,ce}$ corresponds to the donor mass when the OLOF criterion is satisfied. The orbital energy term $\Delta E_{\text{orb}}(M_{env})$ represents the energy released during the inspiral of the companion from the initial separation $a_{OLOF}$ down to a separation equal to the radial coordinate $r(M_{env})$ in the donor-star model outside which an envelope mass $M_{env}$ is located.
We assumed the orbital energy released between the moment of OLOF and the companion entering the donor envelope (when $a = R_{d,ce}$) is also available for envelope ejection.

The envelope binding energy $E_{\mathrm{bind}}(M_{\mathrm{env}})$ is defined as the integral of the gravitational potential energy and internal energy \cite[][]{1994MNRAS.270..121H} (including both thermal and recombination energy) over the outer layers of the donor star, extending from $M_{d,\mathrm{ce}} - M_{\mathrm{env}}$ to the stellar surface at $M_{d,\mathrm{ce}}$.

The term $E_{ej}$ is taken to represent the bulk kinetic energy of the ejecta at infinity. We evaluated it using the ejecta velocity $v_{ej}$, derived from the FWHM of the H$\alpha$ line in the spectrum of each transient (see \citet{Cai2022AA,2023A&A...671A.158P} and Paper II). Specifically, we adopted $v_{ej}=430\pm90\,\mathrm{km\,s^{-1}}$ for AT2021biy and $v_{ej}=460\pm15\,\mathrm{km\,s^{-1}}$ for AT2021blu, accounting for the 1$\sigma$ uncertainties. The impact of these uncertainties on the inferred mass and the expression for the bulk kinetic energy of the ejecta will be discussed in Sect. \ref{sec:discussion}. 

Based on the standard formalism provided in Eqs. \ref{eq:energy_cons} and \ref{eq:energy_terms}, the energy budget can be evaluated at each mass coordinate and the lower-limit estimate is defined as
\[
M_{\mathrm{ej}}^{\mathrm{low}} =
\max_{M_{\mathrm{env}}}
\left\{
M_{\mathrm{env}} \;\middle|\;
\Delta E_{\mathrm{orb}}(M_{\mathrm{env}})
\geq
E_{\mathrm{ej}}(M_{\mathrm{env}}) + E_{\mathrm{bind}}(M_{\mathrm{env}})
\right\},
\]
while the upper-limit estimate is given by
\[
M_{\mathrm{ej}}^{\mathrm{up}} =
\max_{M_{\mathrm{env}}}
\left\{
M_{\mathrm{env}} \;\middle|\;
\Delta E_{\mathrm{orb}}^{*}
\geq
E_{\mathrm{ej}}(M_{\mathrm{env}}) + E_{\mathrm{bind}}(M_{\mathrm{env}})
\right\},
\]
where $\Delta E_{\mathrm{orb}}^{*}$ is the total orbital energy released to the point where the donor’s core fills its Roche lobe. We defined this final orbital separation as the onset of the merger between the companion and the helium core. In Eq. \ref{eq:E_orb}, this corresponds to replacing $r(M_{env})$ by $a_{f}^{*}$, where 
\begin{equation}
\label{eq:final_orbit}
a_{f}^{*} = \max_{a_{f}}
\left\{a_{f}  \;\middle|\; R_{L_{1}}^{\rm core}\left(q^{\rm core}, a_{f}\right) \le R_{\rm d}^{\rm core}\right\}.
\end{equation}
Here $q^{\rm core}=M_{He}/M_{a}$, where $M_{He}$ and $R_{\rm d}^{\rm core}$ are the helium-core mass and radius, respectively, and $R_{L_{1}}^{\rm core}$ is the Roche-lobe radius of the helium core computed using Eq. \ref{eq:RL}.

Our formalism differs from that of \citet{Blagorodnova2021}, particularly in Eq.~\ref{eq:E_orb}, where we assumed that orbital energy deposition begins at a separation equal to $a_{\rm OLOF}$ rather than at the donor surface, and in Eq.~\ref{eq:bulk_kinetic}, where the bulk kinetic energy of the ejecta is evaluated at infinity, such that the gravitational potential term of the ejecta vanishes identically. We also derived an upper-limit estimate of the ejecta mass, whereas \citet{Blagorodnova2021} considered only the lower limit.

\begin{figure}
    \centering
    \includegraphics[width=\linewidth]{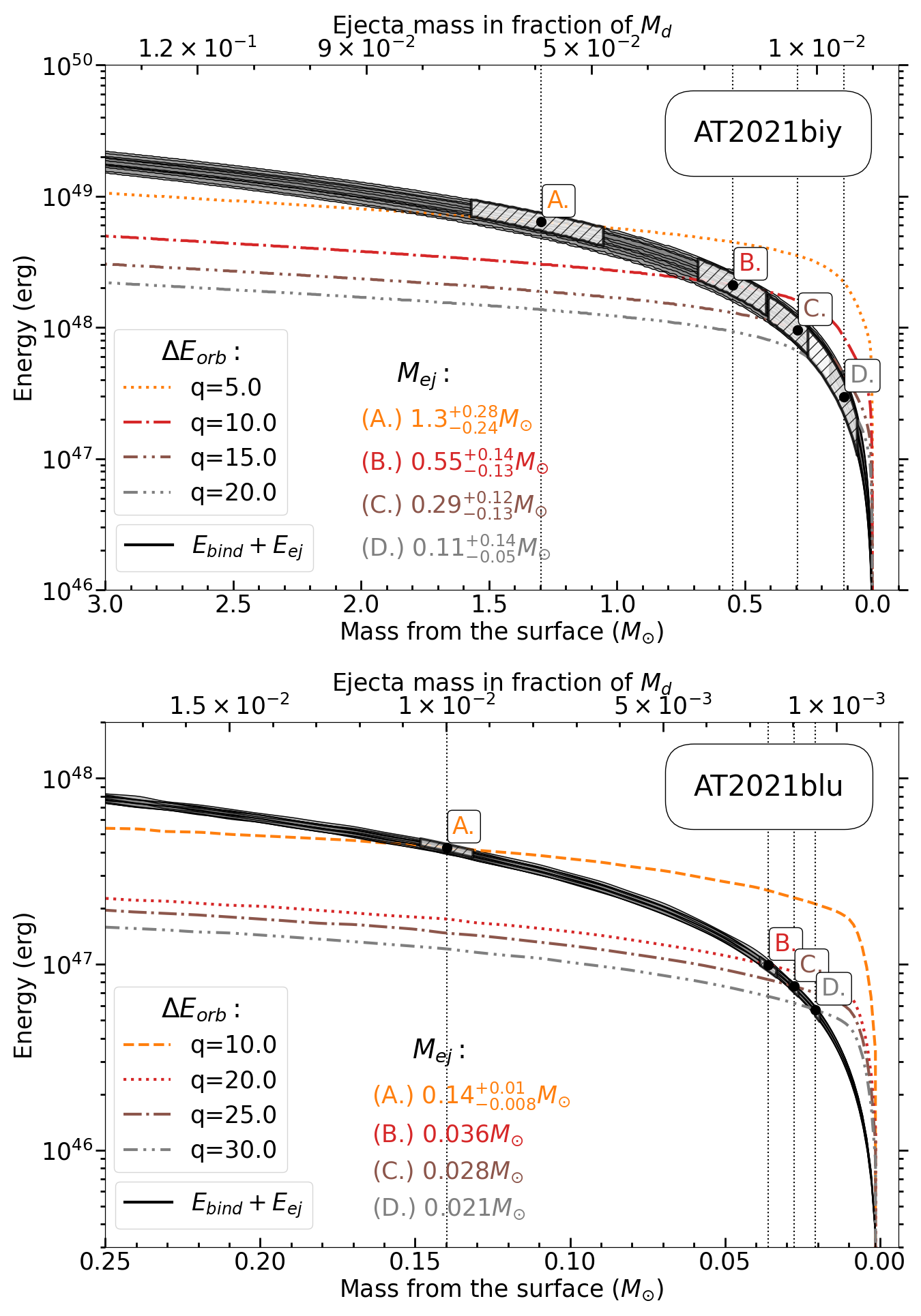}
    \caption{The ejecta mass $M_{ej}$ inferred from the energy-budget at the moment of the OLOF for each mass ratio (shown by intersections A,B,C and D). We also display with a white hatched region, the uncertainties associated with this resolution method. The thickness of the black curve corresponding to the envelope binding energy and the bulk kinetic energy of the ejecta $E_{\rm bind}+E_{\rm ej}$, represents the uncertainties on the observed ejecta velocity ($\pm 90\mathrm{km\,s^{-1}}$ for AT2021biy and $\pm 15\mathrm{km\,s^{-1}}$ for AT2021blu). For AT2021biy, we used MESA binary models with a donor mass $M_{d}=22 M_{\odot}$ and mass ratios $q \in [5,20]$ to compare the energy required to fully unbind an ejected envelope mass $M_{ej}$ with the available energy from companions of different masses. For AT2021blu, we performed the same analysis using progenitor models with $M_{d}=14 M_{\odot}$ and $q \in [10,30]$.}
    \label{fig:AT2021biy_energy}
\end{figure}

Fig. \ref{fig:AT2021biy_energy} illustrates the method in the particular case of a donor mass $M_{d}=22 M_{\odot}$ and mass ratios $q \in [5,20]$ for AT2021biy and $M_{d}=14 M_{\odot}$ and $q \in [10,30]$ for AT2021blu. The ejecta mass estimates correspond to the intersection between the curve representing the orbital energy released on the one hand, and the envelope binding energy plus the bulk kinetic energy of the ejecta on the other. For the full set of models that satisfy the constraints discussed in section \ref{sec:modeling}, we obtained lower limits on the ejecta mass in the range 0.03-2.98 $M_{\odot}$ for AT2021biy and 0.02-0.14 $M_{\odot}$ for AT2021blu. The corresponding upper limits (illustrated in Fig. \ref{fig:upperlimit} in the Appendix) span a broader range, from 1.72-7.82 $M_{\odot}$ for AT2021biy and 1.8-3.8 $M_{\odot}$ for AT2021blu. Considering both bounds, we found that $M_{ej}$ decreases for lower donor masses and higher mass ratios (see Fig. \ref{fig:AT2021biy_Mej_grid} in the Appendix).  We also found that recombination energy accounts for only a small fraction of the total binding energy of the ejected envelope, ranging from 0.22$\%$ to 1.12$\%$ for AT2021biy, and 0.65$\%$ to 1.08$\%$ for AT2021blu. These values are consistent with expectations for massive-star envelopes, in which recombination energy generally represents only a minor contribution to the total binding-energy budget \citep[see, e.g.][]{2016A&A...596A..58K}. 

\subsection{Ejecta mass from light-curve based models}
\label{sec:LC_models}

To place our inferred ejecta masses in context, we also compared the envelope ejected masses derived from this MESA-based energy formalism with predictions from light-curve based models, thereby linking the dynamical and remnant phase to the progenitor properties. In the literature, many studies have attempted to model the observed light curves of LRNe. They initially relied on simple models based on thermal cooling processes \citep[e.g.][]{2013Sci...339..433I,2017MNRAS.470.2339L}, and subsequently added more physical ingedients to address three main mechanisms: hydrogen recombination, shock interaction and dust formation. Recombination-driven models \citep{Matsumoto2022,2024ApJ...963L..35C} suggest that the extended plateau observed in many LRNe, which is particularly long in the case of AT2021biy, is powered by a hydrogen recombination front that recedes through the outflowing mass by releasing stored ionization energy and maintaining a constant luminosity. On the other hand, studies addressing shock interaction \citep{2017ApJ...850...59P,Kirilov2025} argue that the double-peak morphology is due to shocks between the fast merger ejecta and pre-existing slow-moving material resulting from pre-dynamical mass loss, which produces the second luminosity peak. Finally, recent works have investigated the role of dust and geometry \citep[e.g.][]{Hatfull2024} in shaping the infrared evolution. They propose that as the outer ejecta cool down, a mechanism of dust grain nucleation increases opacity and shifts the emission into the infrared.

We estimated the ejecta mass using several analytic and semi-analytic light-curve models developed for Type II-P supernovae and LRNe. These models connect observable properties—such as plateau luminosity, duration, and ejecta velocity\footnote{The velocities used correspond to the same estimates adopted in the energy-budget calculation in Sect. \ref{sec:energetics}.}—to the ejecta mass through different physical assumptions, including recombination-powered diffusion and shock-powered emission. A detailed description of each model and the corresponding estimates is provided in Appendix \ref{app:LC_based_models}. For the observable properties, we adopted $v_{ej}=430\mathrm{km\,s^{-1}}$, $L_{pl} = 5\times10^{40}$ erg.s$^{-1}$, $t_{pl}=210\pm10$d for AT2021biy and $v_{ej}=460\mathrm{km\,s^{-1}}$, $L_{pl} = 3.1\times10^{40}$ erg.s$^{-1}$, $t_{pl}=60\pm20$d for AT2021blu. Depending on the model considered, we found ejected mass values between $0.5M_{\odot}<M_{ej}<19.82M_{\odot}$ for AT2021biy and between $9\times10^{-4}M_{\odot}<M_{ej}<0.55M_{\odot}$ for AT2021blu (see Table \ref{tab:ejecta_mass}).

The dominant source of uncertainty in the input parameters arises from the ejecta velocity. It is therefore instructive to assess each model’s sensitivity to this quantity. To do so, we computed the elasticity $\mathcal{E}_{v_{ej}}=\mathrm{d}\ln M_{\rm ej}/\mathrm{d}\ln v_{\rm ej}$, which is a dimensionless quantity measuring the percentage change in $M_{\rm ej}$ per percentage change in $v_{\rm ej}$. Among the models considered, the Kasen \& Woosley model is the most sensitive (a 1$\%$ increase in the velocity yields a 5.5$\%$ increase in the ejecta mass), whereas the Metzger \& Pejcha model is the least sensitive (with $\mathcal{E}_{v_{ej}}=1$). In addition, because AT2021biy exhibits an unusually long, flat plateau compared to most LRNe, we also quantified each model’s sensitivity to the plateau duration via a similarly defined elasticity parameter $\mathcal{E}_{t_{p}}$. The Kasen \& Woosley model is also the most sensitive to the plateau duration, while Metzger \& Pejcha is the least sensitive. We finally quantified each model’s sensitivity to the plateau luminosity, to facilitate comparisons among the transients. Although AT2021biy is slightly more luminous on the plateau, the generally low sensitivity with respect to $L_p$ in most prescriptions implies only a modest impact on the inferred ejecta mass. 

We compared the predictions of these models for AT2021biy, AT2021blu, and the well-studied LRN AT2018bwo. The latter, located in the NGC 45 host galaxy, exhibits an optical plateau of $\sim41$ days and a cool photosphere ($\sim$ 3300 K). Pre-outburst HST/Spitzer data and binary-evolution modeling point to a binary yellow-supergiant progenitor with primary $\sim$12–16 $M_{\odot}$ undergoing mass transfer and a CE ejection, with an estimated 0.15–0.5 $M_{\odot}$ expelled at $\sim$500 $\mathrm{km\,s^{-1}}$ during the merger \citep{Blagorodnova2021}. Comparing the predictions of the various light-curve based theoretical models, we found that the estimates of the ejecta mass for AT2021blu are similar to those of AT2018bwo. 

\begin{table*}[t]
\centering
\caption{Comparison of physical parameters and mass estimates.} 
\label{tab:comparison}
\renewcommand{\arraystretch}{1.4}
\begin{tabular}{l c c c c c c}
\hline\hline
Parameters/models &  &  &  & AT2021biy & AT2021blu & AT2018bwo\\
\hline
$L_{p}$ ($10^{40}$\,erg/s) &  &  &  & $5 \pm 0.2$ & $3.1 \pm 0.2$ & $1 \pm 0.15$ \\
$t_{p}$ (days)             &  &  &  & $210 \pm 10$ & $60 \pm 20$ & $70 \pm 5$ \\
$v_{ej}$ ($\mathrm{km\,s^{-1}}$) &  &  &  & $430 \pm 90$ & $460 \pm 90$ & $500 \pm 65$ \\ 
\hline
 & $\mathcal{E}_{t_{p}}$ & $\mathcal{E}_{L_{p}}$ & $\mathcal{E}_{v_{ej}}$ & \multicolumn{3}{c}{Ejected Mass $M_{\mathrm{ej}}$ ($M_\odot$)\textsuperscript{a}} \\  
\hline
MESA: envelope mass at OLOF & & & & $10-13.6$ & $9.8-10.1$ & \dots \\
MESA: $M_{ej}^{low}$ (lower-limit estimates)      &  &  & 10\textsuperscript{b} & $0.03-2.98$ & $0.02-0.1$ & $0.02-0.7$ \\
MESA: $M_{ej}^{up}$ (upper-limit estimates)      &  &  & 2.5\textsuperscript{b} & $1.72-7.82$ & $1.8-3.8$ & \dots\\
\cite{2016ApJ...821...38S} & 4 & $-1$ & $-5$ & $19.82^{+9.5}_{-9.8}$ & $\mathbf{0.28^{+0.17}_{-0.13}}$ & $2.06^{+1.05}_{-0.88}$ \\
\cite{2022ApJ...938....5M} & 3 &      & 3    & $9.58^{+4.16}_{-4.79}$ & $\mathbf{0.29^{+0.15}_{-0.13}}$ & $\mathbf{0.55^{+0.31}_{-0.18}}$ \\
\cite{2017MNRAS.471.3200M} & 2 &      & 1    & $\mathbf{6.25 \pm 0.25}$ & $\mathbf{0.55^{+0.09}_{-0.08}}$ & $0.8^{+0.12}_{-0.1}$ \\
\cite{2017ApJ...835..282M} & 4 & $-1$ & $-5$ & $\mathbf{3.92^{+1.88}_{-1.94}}$ & $\mathbf{0.06 \pm 0.03}$ & $\mathbf{0.41^{+0.21}_{-0.17}}$ \\
\cite{2010ApJ...717..245K} & 6 & $-1.5$ & $5.5$ & $\mathbf{0.51^{+0.49}_{-0.37}}$ & $0.0009^{+0.0012}_{-0.0006}$ & $\mathbf{0.02^{+0.02}_{-0.01}}$ \\ 
\hline
 & & & & \multicolumn{3}{c}{Progenitor mass $M_{\mathrm{prog}}$ ($M_\odot$)} \\ 
\hline
This work &  &  &  & $18-23$ & $14 \pm 0.5$ & \dots \\
\cite{Blagorodnova2021}  &  &  &  & \dots & \dots & $12-16$ \\
\hline\hline
\end{tabular}
\caption{Ejected envelope mass ranges for AT2021biy and AT2021blu, based on the lower and upper limits derived from the energy-budget formalism detailed in Sect. \ref{sec:energetics}. These values are compared to the predictions of light curves based models and the case of AT2018bwo \citep{Blagorodnova2021}. Bold values indicate predicted ejecta masses consistent with our MESA models. We also show, each model’s elasticity ($\mathcal{E}_{x}=d\ln M_{ej}/d\ln x$) with respect to plateau duration $x=t_{p}$, luminosity $x=L_{p}$, and ejecta velocity
$x=v_{ej}$, to quantify how sensitive the ejecta-mass estimate is regarding to these parameters. 
\tablefoottext{a}{For AT2021biy, the ejected masses were derived for $M_{d}=18-23M_\odot$ and $q=3-40$, for AT2018blu, they correspond to $M_{d}=14 \pm 0.5 M_\odot$ and $q=10-30$, and for AT2018bwo, the values have been obtained for $M_{d}=15M_\odot$ donor and $q=3-20$.}
\tablefoottext{b}{This elasticity was assessed using a progenitor model with $M_{d}=22M_\odot$,  $q=5$ and $a=700R_\odot$.}}
\label{tab:ejecta_mass}
\end{table*}

The light-curve based models systematically predict larger ejected masses for AT2021biy than for AT2021blu. This mainly reflects the much longer plateau of AT2021biy (by a factor of $\sim 4$), to which most models are strongly sensitive. The longer plateau observed for AT2021biy may reflect a more efficient conversion of kinetic energy into radiation, for instance through stronger shocks and/or a more extended recombination-powered phase. Alternatively, the shorter optical plateau of AT2021blu could be partially attributed to circumstellar obscuration, where newly formed dust absorbs optical photons and re-emits them at infrared wavelengths, effectively suppressing the optical light curve while maintaining the bolometric output (see Sect. \ref{sec:dusty_models}). By contrast, the inferred ejecta velocities are comparable for the two events and therefore play a secondary role in setting $M_{\rm ej}$ in this comparison. Among the light-curve models considered, MacLeod et al. and Kasen \& Bildsten yield the lowest $M_{\rm ej}$ estimates for all three transients and seem more in agreement with the MESA predictions.

When both the lower and upper estimates are considered, the ejecta-mass ranges inferred from our energy-budget approach are broadly consistent with the predictions of various light-curve models. (see Table \ref{tab:ejecta_mass}). While some progenitor models with lower mass ratios ($q\lesssim2$) predict larger ejecta masses, these configurations are likely to lie in the stable mass-transfer regime, rather than in the unstable regime required to initiate a CE phase.


\section{Post-outburst dust evolution}
\label{sec:dusty_models}

Another way to connect the dynamical and remnant phases is to compare our predicted ejected envelope masses with the observed dust mass formed in the remnant. To investigate the post-outburst dust evolution, we used the mid-IR (MIR) fluxes, at W1 (2.75$-$3.87$\mu$m; $\lambda_{\mathrm{c}}$=$3.35\mu$m) and W2 (3.96$-$5.34$\mu$m; $\lambda_{\mathrm{c}}$=$4.60\mu$m) bands.

The analysis of the SEDs of AT2021biy and AT2021blu was carried out using the radiative transfer code {\sc dusty} \citep[v4;][]{1997MNRAS.287..799I} by interpolating the optical SEDs (for the overlapping period), as obtained from \citet{Cai2022AA} and \citet{2023A&A...671A.158P}, respectively, to the times of the NIR observations. For AT2021blu and AT2021biy, we
used ZTF and ATLAS upper-limits, respectively, for the latests epochs.
The post-outburst evolution was modeled following the methodology presented in \citet{2020MNRAS.496.5503B} for M31-LRN-2015.
Using our {\sc pydusty} package\footnote{\url{https://github.com/mgomezAstro/pyDusty}} -- a Python interface for {\sc dusty} --along with the MCMC sampler \citep[{\sc emcee}; ][]{Foreman-Mackey2013PASP}, we derived the best-fit parameters and their uncertainties.
The model assumed a uniform prior on the central source temperature (2000$\leq T_\mathrm{*} \leq$15000\,K, assuming blackbody radiation) and a steady-state wind with an $r^{-2}$ density profile. The dust shell was assumed to have a thickness\footnote{We adopted a thickness of $Y$=2 as our models are insensitive to geometric variations within the limited observational constraints. The available photometry (spanning $\sim$0.3--4.6$\mu$m) lacks coverage beyond the mid-IR, where longer-wavelength emission would probe cooler dust at larger radii and better constrain the shell's outer extent.} $Y$=2, where $Y \equiv R_\mathrm{out}/R_\mathrm{in}$ (with $R_\mathrm{in}$ and $R_\mathrm{out}$ being the inner and outer shell radius, respectively), and was composed of silicate grains \citep{1984ApJ...285...89D} following a canonical ISM size distribution ($a_\mathrm{min}=0.005\mu$m, $a_\mathrm{max}=0.25\mu$m,  and $n(a) \propto a^{-3.5}$).
The visual optical depth, $\tau_\mathrm{V}$, was varied between 0 and 200 and we used the colour excess presented in Sect. \ref{sec:metallicity}.
Finally, we convolved the {\sc dusty} spectrum with the corresponding filter transmission curves, as obtained from the Spanish Virtual Observatory\footnote{\url{https://svo2.cab.inta-csic.es/theory/fps/}}, and compared the resulting synthetic photometry to the observations.
The best-fit models are shown in Figs.~\ref{fig:dusty_models_blu} and \ref{fig:dusty_models_biy}, with parameters and uncertainties listed in Table~\ref{tab:dusty_models}. 

\begin{figure}
    \centering
    \includegraphics[width=\linewidth]{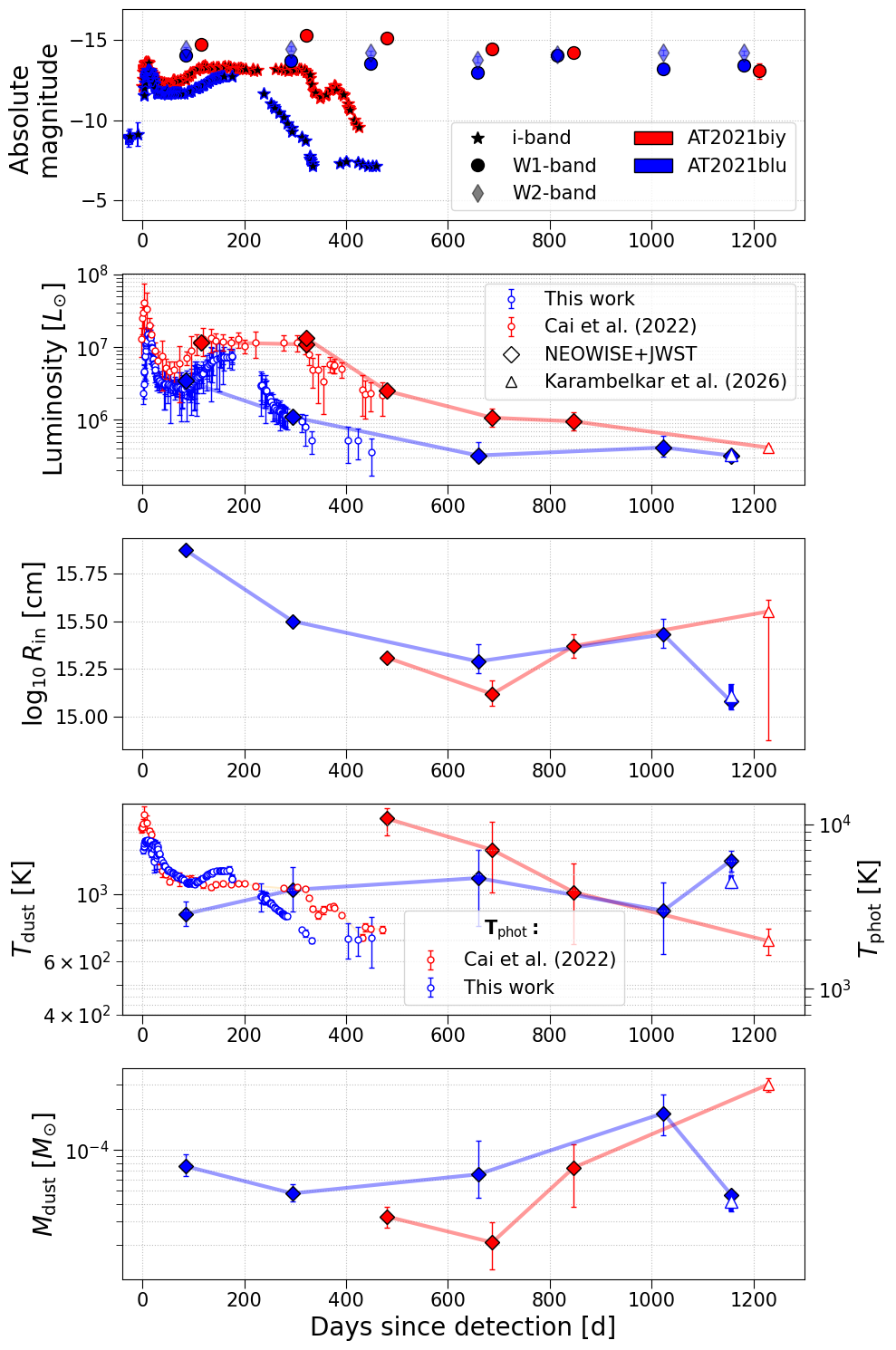}
    \caption{Evolution of the AT2021biy and AT2021blu dust properties. The data displayed here are those presented in Table \ref{tab:dusty_models}. The JWST observations is provided in \citet{2026ApJ...999...16K}. The reference epochs are the discovery dates of each transient, MJD 59243.56 for AT2021biy and MJD 59246.467 for AT2021blu.}
    \label{fig:dust_evol}
\end{figure}

To calculate the dust masses of AT2021blu and AT2021biy, we followed the formulation derived by \citet{2025ApJ...983...87L} as,
\begin{equation}
\label{eq:dust_mass}
    \frac{M_\mathrm{dust}}{M_\sun} \approx 2.4\times 10^{-11} \left( \frac{R_\mathrm{in}}{50R_\sun} \right)^{2} \left( \frac{\tau _\mathrm{V}}{0.7} \right) \left( \frac{Y}{5} \right) \left( \frac{1.123 \times 10^{4}\text{cm}^{2}\text{g}^{-1}}{\kappa _\mathrm{V}} \right),
\end{equation}
where $\kappa _\mathrm{V}$ is the dust opacity per mass of dust, and $R_\mathrm{in}$, $\tau _\mathrm{V}$, and $Y$ are the previously described parameters from the {\sc dusty} models.
Here, we assumed $\kappa _\mathrm{V}=1.123\times 10^{4}\text{cm}^{2}\text{g}^{-1}$ which is consistent with the optical properties from \citet{1984ApJ...285...89D}.
The calculated dust masses, dust temperature, and radius for the different epochs of both LRNe are summarized in Table~\ref{tab:dusty_models}. The evolution of the luminosity and the dust properties are shown in Fig. \ref{fig:dust_evol}, which also combines the recent analysis of  \cite{2026ApJ...999...16K} using JWST data.

\subsection{AT2021biy}

\begin{figure*}
	\centering
	\includegraphics[width=1.0\textwidth]{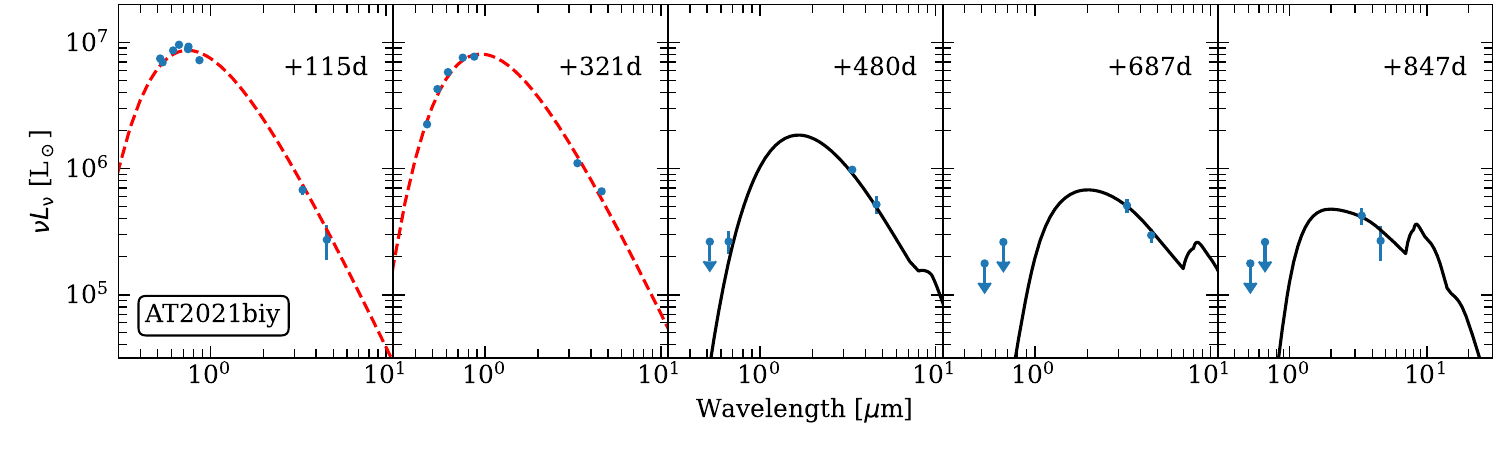}
	\caption{Similar to \ref{fig:dusty_models_blu} but for AT2021biy phases measured to the MJD 59243.56 and a correction for foreground reddening of $E(B-V)$=0.271mag. \label{fig:dusty_models_biy}}
\end{figure*}

In AT2021biy, the earliest epoch with full optical–NIR SED coverage occurs at 115\,d, corresponding to the onset of the second peak in the $r$-band. At this epoch, the SED is well reproduced by a single blackbody with a temperature of $T_{\mathrm{*}} \sim 4945$\,K and a radius of $4651R_{\odot}$ (red dashed line, first panel of Fig.~\ref{fig:dusty_models_biy}).
This is consistent with a photospheric radius expected for constant velocity expansion at $\sim430\mathrm{km\,s^{-1}}$ \citep[see][]{Cai2022AA}. 
For the second epoch at 321\,d, a single blackbody still provides a satisfactory fit, yielding a radius of $\sim$7204\,$R_{\odot}$, significantly smaller than the expected photospheric radius of 17\,142\,$R_{\odot}$ at such expansion velocity (red dashed line, second panel of Fig.~\ref{fig:dusty_models_biy}). The absence of an infrared excess, combined with the relatively small inferred blackbody radius, suggests that the ejecta have not yet interacted with any pre-existing dust. which would be too cold to be detectable in the progenitor phase.
However, the subsequent evolution requires the inclusion of a dust component to reproduce the SED and reveals a progressive increase in optical depth together with a gradual decrease in dust temperature, while the inferred inner dust radius remains approximately constant. Between 480\,d and 687\,d, the uncertainties are too large to robustly establish a slight decline in the dust mass. We can only confidently infer an overall increase from 480\,d to 1228\,d. This evolution is consistent with that of the dust temperature, which varies in the opposite direction.

The high dust temperature of 1775\,K at 480\,d may indicate that the MIR emission arises from pre-existing dust that is radiatively heated by ongoing circumstellar interaction between the ejecta and the surrounding material. 
However, such a high temperature exceeds the typical sublimation temperature of silicate dust \citep[$\sim$1500\,K; e.g.][]{Draine2011}, suggesting that silicate grains would rapidly evaporate under these conditions.
Alternatively, the elevated dust temperature may result from high metal vapour pressure in the dust-forming region, which can increase the temperature at the inner boundary of the dust shell without requiring the presence of carbonaceous dust \citep{2012ApJ...760..123R}. As the interaction weakens, the dust temperature decreases, reaching $\sim$700\,K by 1228\,d.
In this context, the detection of Fe\,{\sc ii} forest lines in the early-time spectra of AT2021biy \citep[see][and Paper II]{Cai2022AA} may indicate the possibly presence of gaseous iron in the environment of the transient, qualitatively consistent with a metal-rich circumstellar region.
Nevertheless, the derived dust properties should be treated with caution. Owing to the position of the source within the host galaxy and the relatively low spatial resolution of WISE, the MIR photometry may be contaminated by background emission from the host galaxy, potentially affecting the SED fitting and the inferred dust parameters.

\subsection{AT2021blu} 

\begin{figure*}
	\centering
	\includegraphics[width=1.0\textwidth]{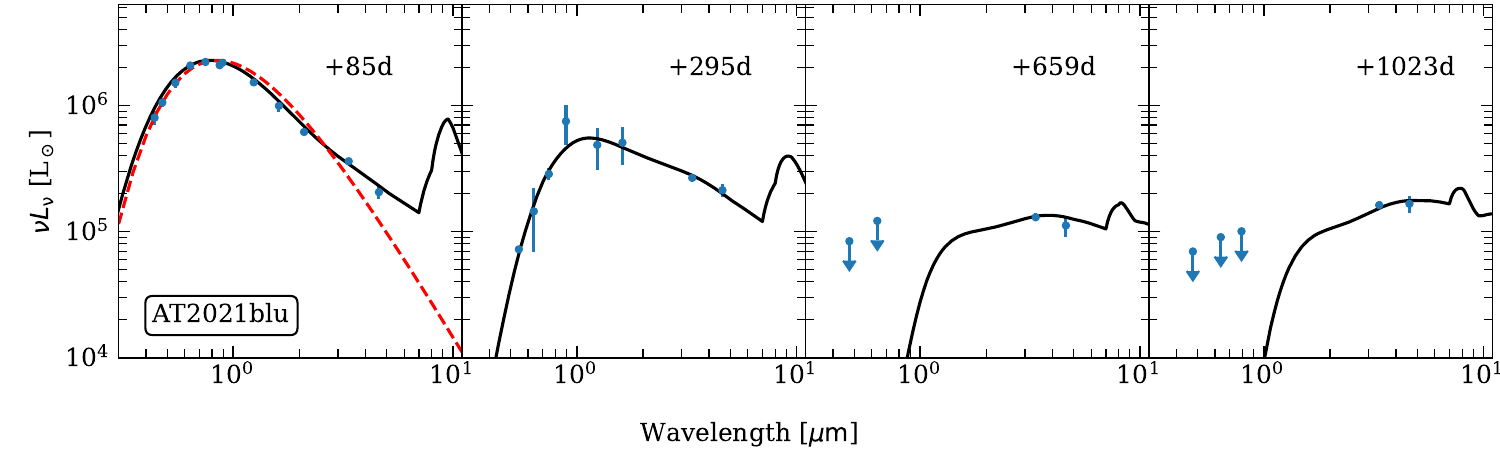}
	\caption{Evolution of the SED of AT2021blu with phases measured relative to the MJD 59246.467. Observed fluxes were corrected for foreground reddening of $E(B-V)$=0.02mag. The black solid line shows the best-fitting {\sc dusty} model. A blackbody is shown (red dashed line) for reference at the first epoch. \label{fig:dusty_models_blu}}
\end{figure*}

Our earliest full optical-to-NIR SED for AT2021blu corresponds to the onset of the second peak in the $r$-band at 85\,d.
At this phase, a single blackbody fit (red dashed line, left panel of Fig.~\ref{fig:dusty_models_blu}) yields a stellar temperature $T_\mathrm{*} \sim 4349$K, $\log(L/L_{\sun})\sim$ 6.49, and a stellar radius, $R_{*}\sim$ 3073R$_{\sun}$, lower than the expected photospheric size of $~4800$R$_{\sun}$ given the high-velocity component at this stage \cite[$\sim$ 460$\mathrm{km\,s^{-1}}$;][]{2023A&A...671A.158P}.
However, the SED shows a small MIR excess, indicating the possible presence of dust. Using our dusty modelling, we derive that the best-fit model requires an additional optically thin dust component, situated at $\sim$495\,au from the central source (see Table~\ref{tab:dusty_models}).
At such a distance, the material would need a velocity of $\sim 10^{4}\mathrm{km\,s^{-1}}$—exceeding typical LRN velocities but similar to those seen in core-collapse SN ejecta. 
Instead, we favoured a scenario in which this shell consists of pre-existing dust, formed in the years preceding the outburst.
This interpretation is supported by the derived dust temperature of $T_\mathrm{dust} \sim 859$K, 
consistent with the modelling presented for AT2021blu's progenitor in Sect. \ref{sec: archival_data}, which showed evidence of a 1000\,K dust emission component appearing in 2012 (see Fig.~\ref{fig:SED_evol}).

During subsequent evolution of AT2021blu through the plateau phase and $r$-band decline (85–659\,d), the bolometric luminosity declines from $\log L/L_{\sun} = 6.54$ to 5.51, while $T_{\star}$ decreases from $\sim 5012\mathrm{K}$ to $\sim 3705\mathrm{K}$. Concurrently, the dust component exhibits an increasing optical depth ($\tau_{\mathrm{V}}$), slight variations in the dust temperature ($T_{\mathrm{dust}}$), and a decreasing inner radius ($\log R_{\mathrm{in}}$). These trends suggest ongoing dust processing in the vicinity of a pre-existing dust component, possibly within the interaction region bounded by the forward and reverse shocks generated by the collision between the dynamically ejected material and an earlier phase of equatorially concentrated mass loss \citep[see][]{2013MNRAS.434..102S,2017MNRAS.471.3200M}. 
While the inner dust radius shows no significant evolution between 295 to 1023\,d epochs, the dust mass exhibits a steady increase, from $\log M_\mathrm{dust}/M_\sun = -4.32$ at 85\,d up to $-3.73$ at 1023\,d, implying ongoing dust formation.
Despite that in \citet{2026ApJ...999...16K} we used the latest NEOWISE photometry containing the persistence flag set, which may be affecting the resulted dust temperature and mass, we obtained very similar results by using only the 2MASS and JWST photometry presented there (see Table\,\ref{tab:dusty_models}). 
For this latest epoch, both the dust radius and mass decrease from 180\,au and $-$3.73 to 80\,au and $-$4.33, respectively, which may indicate a dust destruction mechanism.
This behaviour may support the presence of shocked material in the circumstellar environment, as strong radiation fields associated with shock-heated gas can sublimate dust grains at high temperatures. Note that our models show a dust temperature of $T_\mathrm{dust}\sim1300$K at 1156\,d, close to the sublimation temperature of silicate grains.

For both AT2021biy and AT2021blu, the measured dust masses are significantly smaller than the inferred ejected-envelope mass (see Sect. \ref{sec:energetics}), primarily because only a small fraction of the expelled gas is expected to condense into dust grains. Moreover, the limited wavelength coverage of NEOWISE restricts the observations to the warm dust component, leaving colder dust undetected and thus unaccounted for in the mass estimates. In addition, shocks and radiative processes may destroy dust grains, and since dust formation can occur over timescales of decades \citep[e.g.,][]{Bermudez2024}, this would further reduce the inferred dust mass.

\begin{table*}
    \renewcommand{\arraystretch}{1.2} 
	\centering
	\caption{Posterior parameters from {\sc dusty} MCMC models of AT2021blu and AT2021biy remnant SEDs. The reference epochs for computing the phases are the discovery dates of each transient, MJD 59243.56 for AT2021biy and MJD 59246.467 for AT2021blu. \label{tab:dusty_models} }
	\begin{tabular}{cccccccccc}
		\hline \hline
		Date	&	Phase	&	$\log L/L_{\sun}$	&	$T_\mathrm{*}$	& $\log(R_{*})$\tablefootmark{a}	& $\tau_\mathrm{V}$	&	$T_\mathrm{dust}$ & $\log R_\mathrm{in}$	&	$\chi^{2}_\mathrm{min}$ & $\log M_\mathrm{dust}/M_{\sun}$\\
		(UTC)	&	(d)		&					&	(K)				& (cm)	& 					&		(K)	& (cm)	&	&					   \\
		\hline
		\multicolumn{10}{c}{AT2021blu} \\
		\hline
		2021-04-27 &  85 & 6.54$^{+0.02}_{-0.01}$ & 5012$^{+114}_{-109}$ & 14.24$^{+0.01}_{-0.01}$  & 1.19$^{+0.18}_{-0.14}$ & 859$^{+87}_{-75}$ & 15.87$^{+0.01}_{-0.01}$ & 14$^{+3}_{-2}$ & $-$4.12$^{+0.09}_{-0.07}$ \\
        2021-11-23 & 295 & 6.04$^{+0.04}_{-0.03}$ & 4253$^{+416}_{-342}$ & 14.16$^{+0.08}_{-0.05}$ & 4.29$^{+0.32}_{-0.35}$ & 1032$^{+191}_{-156}$& 15.50$^{+0.02}_{-0.02}$ &7$^{+3}_{-2}$ & $-$4.32$^{+0.07}_{-0.06}$ \\
		2022-11-22 & 659 & 5.51$^{+0.18}_{-0.09}$ & 3705$^{+1548}_{-1250}$& 13.98$^{+0.41}_{-0.16}$ & 15.84$^{+7.71}_{-8.79}$ & 1129$^{+269}_{-346}$ & 15.29$^{+0.09}_{-0.06}$ & 2$^{+2}_{-1}$ & $-$4.18$^{+0.25}_{-0.17}$ \\
        2023-11-21 & 1023& 5.62$^{+0.16}_{-0.13}$  & 3284$^{+1047}_{-897}$ & 14.14$^{+0.27}_{-0.13}$ & 23.51$^{+10.83}_{-10.14}$ & 881$^{+214}_{-250}$ & 15.43$^{+0.08}_{-0.07}$ & 2$\pm^{2}_{1}$ & $-$3.73$\pm^{0.14}_{0.16}$ \\
         2024-04-03\tablefootmark{b} & 1156  & 5.52$^{+0.02}_{-0.03}$  & 2200$^{+200}_{-200}$ & 14.44$^{+0.09}_{-0.09}$ & 22.4$^{+1.2}_{-1.3}$ & 1100$^{+50}_{-50}$ & 15.11$^{+0.06}_{-0.07}$ & $\cdots$ & $-$4.38$^{+0.07}_{-0.07}$ \\
		\hline
		\multicolumn{10}{c}{AT2021biy} \\
		\hline
         \textbf{2021-05-24} & 115 & 7.07$^{+0.01}_{-0.01}$ & 4945$^{+30}_{-27}$ & 14.51$^{+0.01}_{-0.01}$ & $\cdots$ &  $\cdots$ &  $\cdots$ &  299$^{+2}_{-1}$ & $\cdots$ \\
         \textbf{2021-12-16} & 321 & 7.04$^{+0.01}_{-0.01}$ & 3885$^{+12}_{-11}$ & 14.70$^{+0.01}_{-0.01}$ & $\cdots$ &  $\cdots$ &  $\cdots$ &  126$^{+2}_{-1}$ & $\cdots$ \\
		2022-05-24 & 480 & 6.40$^{+0.04}_{-0.04}$ & 5437$^{+2600}_{-1562}$& 14.09$^{+0.54}_{-0.20}$ & 6.88$^{+0.67}_{-0.98}$ & 1775$^{+152}_{-217}$ & 15.31$^{+0.02}_{-0.02}$ & 3$^{+3}_{-2}$ & $-$4.49$^{+0.07}_{-0.08}$\\
        2022-12-17 & 687 & 6.03$^{+0.13}_{-0.12}$ & 2796$^{+811}_{-528}$& 14.48$^{+0.23}_{-0.09}$ & 10.70$^{+7.95}_{-5.69}$ & 1399$^{+331}_{-388}$ & 15.12$^{+0.07}_{-0.06}$ & 3$^{+2}_{-1}$ & $-$4.68$^{+0.15}_{-0.20}$\\
        2023-05-26 & 847 & 5.98$^{+0.12}_{-0.12}$ & 2684$^{+1041}_{-757}$& 14.49$^{+0.36}_{-0.16}$ & 12.04$^{+8.21}_{-6.78}$ & 1014$^{+248}_{-332}$ & 15.37$^{+0.06}_{-0.06}$ & 2$^{+2}_{-1}$ & $-$4.13$^{+0.17}_{-0.29}$\\
         2024-06-08\tablefootmark{b} & 1228  & 5.62$^{+0.02}_{-0.02}$  & 2400$^{+600}_{-600}$ & 14.41$^{+0.26}_{-0.20}$ & 20.3$^{+2.8}_{-2.3}$ & 700$^{+70}_{-70}$ & 15.55$^{+0.06}_{-0.67}$ & $\cdots$ & $-$3.52$^{+0.05}_{-0.05}$  \\
		\hline
	\end{tabular}
    \tablefoot{ Bold text in the \textit{Date} column indicates that only a Blackbody was fitted. 
    In this case the model returned the stellar radius, $R_{*}$, and temperature, $T_{*}$, and the luminosity, $\log(L/L_{\sun})$ was calculated with $L=4\pi \sigma_\mathrm{SB} R_{*}^{2} T_{*}^{4}$, where $\sigma_\mathrm{SB}$ is the Stefan-Boltzmann constant.\\
    \tablefoottext{a}{Calculated with $R_{*}=\sqrt{L_{*} / 4\pi\sigma_{SB}T_{*}^{4}}$}.
    \tablefoottext{b}{These data have been obtained from \citet{2026ApJ...999...16K}.}
    }
\end{table*}

\section{Discussion}

We now discuss how uncertainties in the observational constraints affect our results, before turning to the main theoretical limitations of our energy-budget approach for estimating the ejected envelope mass. We also outline ways to mitigate observational uncertainties and propose a methodological refinement for estimating the mass of the ejecta.

\label{sec:discussion}
\begin{figure}
    \centering
    \includegraphics[width=\linewidth]{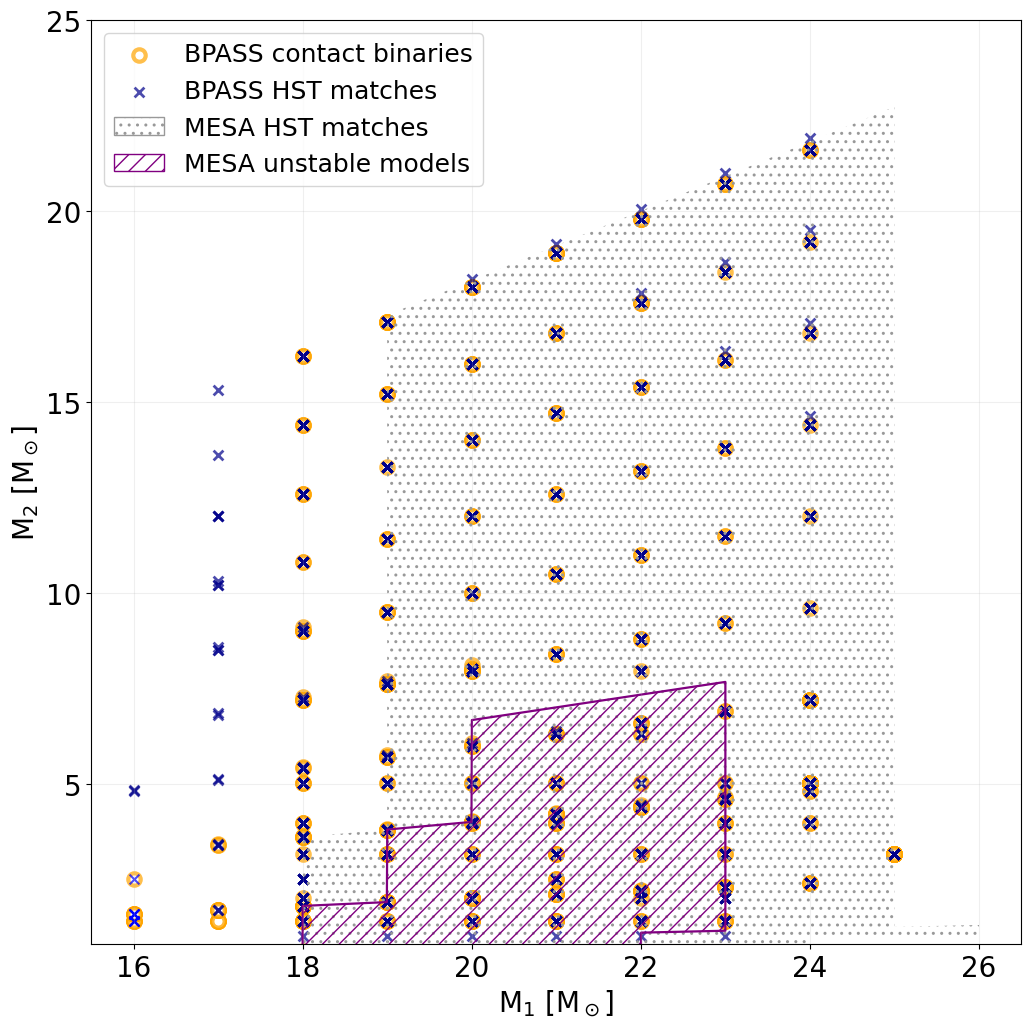}
    \caption{BPASS progenitor systems of AT2021biy, identified by \cite{Cai2022AA} as consistent with the HST observations. Overlaid are the constraints obtained from our MESA models, including the OLOF instability criterion, which further restrict the allowed parameter space to a smaller subregion.}
    \label{fig:BPASS_vs_MESA}
\end{figure}

One major improvement of our analysis compared to the detailed binary modelling presented in \cite{Blagorodnova2021} is that we explicitly required mass transfer to be unstable when the model intersects the observed HR diagram location of the progenitor, ensuring that the mass-transfer rates is sufficiently high to be consistent with the onset of a CE phase. Although it remains challenging to explore the full parameter space of binary progenitors, we have shown that applying mass-transfer instability constraints to binary models helps to narrow down possible binary configurations.

Our method reduces the parameter space of AT2021biy progenitors to $M_{d} = 18-23 M_{\odot}$ with $q\gtrsim3$ (see Fig. \ref{fig:BPASS_vs_MESA}), and to $M_{d} = 14 \pm 0.5 M_{\odot}$ with $q\gtrsim10$ for AT2021blu. These donor-mass estimates are consistent with those reported by \cite{2023A&A...671A.158P}, who derived $M_{d} = 14 \pm 1 M_{\odot}$ for AT2021blu using single-star models and $M_{d} = 13-16 M_{\odot}$ from BPASS binary models. They are also in agreement with \cite{Cai2022AA}, who, using BPASS, found had a primary star mass of $M_{d} = 17-24 M_{\odot}$ for AT2021biy. However, these studies provide only limited constraints on the companion mass: $\sim60\%$ of the BPASS models have mass ratios $q \simeq 3-5$ for AT2021biy in \cite{Cai2022AA} and $q>2$ for AT2021blu \citep{2023A&A...671A.158P}. Moreover, they do not constrain the initial binary separation, which is left as a free parameter. 

This relative agreement with previous studies contrasts with the conclusion of \citet{Blagorodnova2021}, who reported that detailed binary-evolution models for AT2018bwo imply a progenitor mass $9-45\%$ larger than that those inferred from single-star evolutionary models. The agreement we find between single and binary models in our case is likely due to the onset of instability occuring near the beginning of OLOF, shortly after Roche-lobe overflow. At this stage, the luminosity decrease is moderate, and the inferred progenitor properties therefore remain close to those predicted by single-star models.

\subsection{Observational limitations}

To derive observational constraints on our progenitors, we assumed that a single blackbody provides a good approximation of the system’s SED, as the donor is expected to dominate the total flux. We further adopted the metallicity and distance values reported in the literature \citep[see][]{2023A&A...671A.158P, Cai2022AA} as reliable proxies. In the following, we discuss these choices in detail and assess how the associated uncertainties affect our results.

The photometry and blackbody fitting procedures are key steps in our methodology, as they establish the region of the HR diagram in which the different progenitor models must reside while undergoing unstable mass transfer. We assumed that the pre-outburst SED for both events could be described by a single blackbody model, which is a strong assumption. However, the available data do not provide enough information to constrain a more detailed model. We therefore assumed that the flux in the observed bands is dominated by the primary star and that binary interaction produces no additional luminosity sources. While a small, hot companion might contribute to the bluer filters (e.g., F606W), we assumed its intrinsic luminosity (photospheric emission) is negligible compared to the donor. This is equivalent to assuming a high progenitor mass ratio, ($M_{d}/M_{a}$), which is consistent with what we found in Sect. \ref{sec:modeling}. We also disregarded any contribution from shocks and accretion-powered emission during the pre-outburst phase, which could otherwise alter the optical stellar spectrum.

Uncertainty in the local metallicity at the transient locations can significantly affect the modeling. Metallicity is known to fundamentally influence the radial expansion of massive stars, thereby dictating both the timing and the specific regime of mass transfer \citep[see, e.g.,][]{2020A&A...638A..55K}. Both AT2021biy and AT2021blu appear to have LMC-like to sub-LMC metallicities, with $Z_{biy}\approx0.007$ and $Z_{blu}\approx0.005$. Future environment studies at the location of each transient will be needed to provide more accurate metallicities as a starting point for the modeling.

A further limitation is that we did not account for distance uncertainties for the modeling of the progenitor SED. As a first-order approximation, this is justified because reddening errors typically dominate the uncertainty in the estimated blackbody parameters. It can be shown that distance uncertainties mainly act as a pure normalization, rescaling $L$ (and the inferred radius $R$) with little effect on $T_{eff}$. In contrast, reddening uncertainties modify the spectral shape and colors, biasing the inferred temperature $T_{eff}$. For AT2021biy, we performed a sanity check by refitting the SED using $E(B-V)\pm\sigma_{E}$ and $D\pm\sigma_{D}$ separately, in order to quantify their respective effects on the inferred parameters. As expected, distance uncertainties primarily affect the inferred radius $R~(\sim7\%)$ and have a negligible impact on the effective temperature $T_{eff}~(\sim0.3\%)$. In contrast, reddening uncertainties lead to significantly larger variations, of order $\sim14\%$ in radius and $\sim43\%$ in effective temperature. 

The proxy adopted to estimate the ejecta velocity is also subject to significant uncertainty. Following previous studies (see \citet{Cai2022AA,2023A&A...671A.158P} and Paper II), we used the FWHM of the \io{H}{$\alpha$} line profile. While this traces the bulk motion of the gas, it neglects the highest velocities present in the distribution tails, which may be further influenced by geometrical effects such as asphericity and inclination. In addition, the velocity measurements depend sensitively on the quality of the available spectroscopic data, particularly the instrumental resolution, which in some cases is comparable to the measured FWHM. In this regards, the inferred ejecta velocity should be regarded as a crude approximation.

We evaluated the impact of uncertainties in the ejecta velocity on the envelope masses derived from our energy-budget formalism. Increasing the ejecta velocity systematically reduces the inferred ejecta mass in both the lower- and upper-limit cases, with a dependence that strengthens with increasing mass ratio. In the lower-limit regime, as the adopted asymptotic velocity ($v_{\mathrm{ej},\infty} = 430\,\mathrm{km\,s^{-1}}$) significantly exceeds the escape velocity (e.g., $v_{\mathrm{esc}} \approx 176\,\mathrm{km\,s^{-1}}$ for $M_d \approx 20\,M_{\odot}$ and $R_d \approx 241\,R_{\odot}$), the energy budget is dominated by the bulk kinetic energy of the ejecta. In particular, we found that for a 1$\%$ increase in $v_{ej,\infty}$, the fractional change in the ejecta mass estimate approximately scales as $\Delta M_{ej} / M_{ej} \approx-0.03\times q^{0.74}$. For example, at $q=5$, this corresponds to a reduction of $\sim10\%$ in the inferred ejecta mass. Since the actual uncertainties on ejecta velocities are expected to be significantly larger than $1\%$, this illustrates the strong sensitivity of the energy-budget estimates to both velocity uncertainties and the binary mass ratio. By contrast, the upper-limit estimate is primarily governed by the envelope binding energy and is therefore much less sensitive to variations in the ejecta velocity. In this case, the dependence can be approximated as $\Delta M_{ej} / M_{ej}\approx-0.01\times q^{0.56}$, corresponding to a reduction of  $\sim2.5\%$ at $q=5$. Most light-curve-based models also exhibit sensitivity to velocity uncertainties, as shown in Table~\ref{tab:ejecta_mass}. As a result, our ejecta-mass estimates, particularly the lower-limit ones, should be interpreted with caution.

\subsection{Theoretical limitations affecting the ejecta mass estimate}

In this section, we distinguish between the assumptions made in the progenitor modeling and those adopted in our energy-budget formalism. We then assess how each of these affects the overall uncertainties in our modeling and ejecta-mass estimates, and outline possible improvements or alternative approaches.

\subsubsection{Binary system modeling}

An important limitation of our approach is the assumption of spherical symmetry throughout the dynamics. This assumption is implicitly adopted in all the methods explored in this study and prevents us from accounting for geometry-dependent effects arising in asymmetric outflows, which are likely prevalent in LRNe \citep[e.g.,][]{2012ApJ...746...74R,Kirilov2025}. In such cases, predictions based on one-dimensional models may become unreliable. In particular, this assumption is inherent to the use of one-dimensional stellar evolution codes such as MESA. By construction, the reliability of MESA predictions decreases as the system becomes increasingly aspherical due to Roche geometry, tidal distortions, or mass transfer. Such asymmetries become even more important at the onset of OLOF, and therefore relying too heavily on MESA outputs beyond this phase may be problematic. A natural extension would be to incorporate insights from three-dimensional simulations to calibrate or correct the MESA-based calculations \citep[e.g.,][]{2025A&A...702A..61R}.

We adopted a simplified description of the companion as a point mass, such that the binary interaction is reduced to a gravitational potential governing mass transfer from the donor. This approximation introduces an important limitation. Because the companion is not assigned any finite size, internal structure, or evolutionary state, its response to mass accretion is undetermined (e.g.\ the extent to which it develops rotation, or transfers material back onto the donor). In particular, in a merger scenario where the companion spirals inward through the donor’s envelope, the outcome may depend sensitively on the compactness of the companion. If it is less compact than the donor’s helium core and retains an extended envelope (e.g., a MS star), it may fill its own Roche lobe before the core does. In such a case, the merger conditions defined in Sect.~\ref{sec:energetics} no longer apply. A natural improvement would be to model the structure of the companion using MESA as well, although this would significantly increase the complexity of the modeling and its computational cost.

By assuming that half of the transferred mass is lost through isotropic re-emission and half through a circumbinary toroid, we effectively adopted a fully non-conservative mass-transfer scenario. In case the accretor is a compact object, this is consistent with the expectation that accretion is Eddington-limited, which implies very small accretion efficiencies given the high mass-transfer rates we find in our progenitor models. In the case of MS accretors, two mechanisms are commonly invoked to limit accretion and thereby favor non-conservative mass transfer. The first is thermally limited accretion, in which the accretor cannot accept mass at a rate far exceeding its thermal timescale without expanding and potentially forming a contact binary \citep{Tout1997,Hurley2002}. In our case, most viable progenitor models involve high mass ratios, implying that any MS accretor has a very long thermal timescale compared to the timescale of mass transfer, thus strongly favoring non-conservative evolution. The second mechanism is rotationally limited accretion, in which spin-up of the accretor to critical rotation prevents further accretion after only a small amount of mass has been transferred \citep{Packet1981,Langer2003}. This picture has recently been challenged by observations of Be+sdOB systems \citep[e.g.,][]{2025ApJ...990L..51L}, which support high mass-transfer efficiencies in stable early Case-B binaries with initial donor masses of $\sim 2-9M_{\odot}$ and mass ratios $<5$. However, the progenitors considered here lie in a significantly more massive regime and, more importantly, have much more extreme mass ratios. They are therefore expected to develop much higher mass-transfer rates, again favoring non-conservative evolution. 

For the lower-mass AT2021blu progenitors, the quasi-adiabatic or dynamical instability criteria are reached closer to the onset of RLOF and are more likely to be satisfied within the observed HR-diagram constraints (see Fig.~\ref{fig:candidates}). In contrast, we did not find AT2021biy progenitors that simultaneously satisfy these criteria and the observational constraints. However, as illustrated in Fig.~\ref{fig:candidates_Mdot}, the onset of OLOF appears to be an appropriate proxy for instability in our selected progenitors, as the mass-transfer rate has already entered a runaway phase. Moreover, for both transients, the quasi-adiabatic and dynamical criteria are generally met only after the onset of OLOF. We therefore adopted the latter as a proxy for the onset of dynamical instability. This approximation should nevertheless be treated with caution. Some systems (e.g., tracks I and II in Figs.~\ref{fig:SE_types}c and \ref{fig:SE_types}d) exhibit a delayed runaway in $\dot{M}_{d}$, with the associated instability developing significantly after OLOF. In such cases, typically corresponding to larger initial binary separations, the delayed onset of runaway mass transfer may invalidate the use of OLOF as a proxy for dynamical instability. A detailed analysis of the mass-transfer rate evolution is therefore required to reliably determine the instability regime experienced by the system.

Another limitation of our approach is the poorly constrained progenitor mass ratio, for which we cannot derive a robust upper bound. As discussed in Sect. \ref{sec:instability}, increasing the mass ratio shifts the onset of OLOF closer to that of RLOF, while also causing the system to approach the dynamical regime more rapidly. Constraining the upper bound depends on the relative position of the onset of RLOF with respect to the observed HR region. If RLOF occurs outside this region, a critical mass ratio exists above which OLOF also falls outside, thereby defining an upper bound on $q$. Conversely, if RLOF already occurs within the observed region, no such upper bound can be derived. Independently, for a given donor-star mass, higher mass ratios lead to lower inferred ejecta masses (Fig. \ref{fig:AT2021biy_Mej_grid}). We can therefore use the remnant modeling of Sect. \ref{sec:dusty_models}, which provides a lower limit on the post-outburst dust mass, to constrain $q$: since the total ejected mass must exceed the observed dust mass, this translates into an effective upper limit on the mass ratio. In practice, this constraint remains well above the range of mass ratios explored in our models.

\subsubsection{Energy-budget formalism}

Regarding our energy-budget methodology, we assumed that energy conversion was fully efficient during the CE phase ($\alpha_{CE}=1$). We recognize that the assumption of perfect energy conversion in the absence of losses is a major limitation of our approach. In practice, radiative cooling, \textit{PdV} expansion work, shocks, and other dissipative processes reduce the fraction of energy available to unbind the envelope, implying that the effective energy budget for ejection is likely lower than assumed here. Allowing for lower common-envelope efficiencies ($\alpha_{CE}<1$) would therefore reduce the inferred ejecta mass. Accordingly, the estimates derived in Sect. \ref{sec:energetics} are better understood as indicative values for the case $\alpha_{CE}=1$, rather than strict lower limits.

In the case of the lower-limit estimate of the ejecta mass, we adopted the strong assumption of local and instantaneous energy deposition, although this is unlikely to be realized in practice. To address this limitation, we developed an improved shell-based formulation (see Sect.~\ref{new_algo} in the Appendix) that allows for non-local, delayed ejection, while retaining the assumption that the companion spirals in until the helium core of the donor fills its Roche lobe. For the illustrative case $M_{d}=22M_\odot$, $q=5$ and $a=700R_{\odot}$, this method yields an ejected mass of $2.1 M_{\odot}$ (see Fig. \ref{fig:AT2021biy_newalgo}), intermediate between the lower and upper estimates of $1.3 M_{\odot}$ and $6.42M_{\odot}$, respectively, derived with our original prescription (see Fig.~\ref{fig:AT2021biy_energy}).

\begin{figure}
    \centering
    \includegraphics[width=\linewidth]{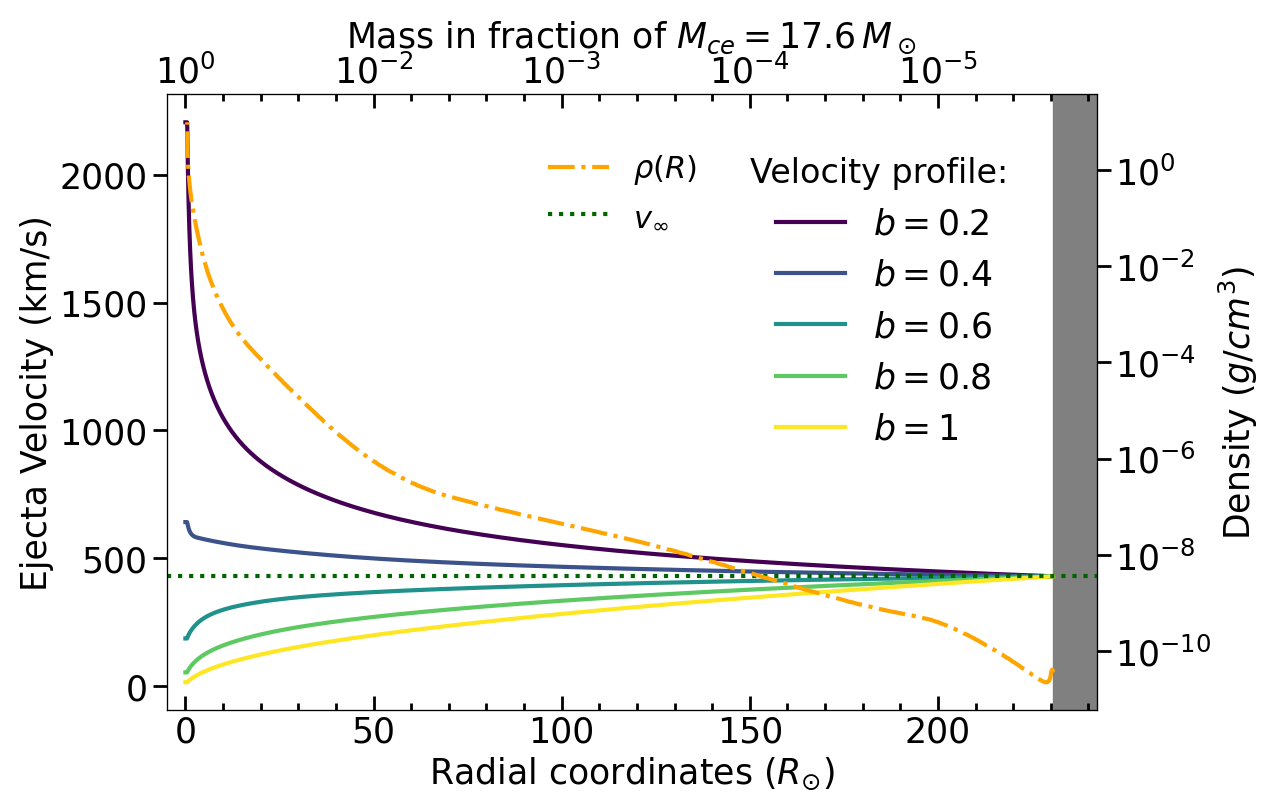}
    \caption{Ejecta velocity as a function of radius for several values of the velocity–profile parameter $b$ in the case of a $M_{d}=22M_{\odot}$ donor star for AT2021biy. As a reference, we also plot the density profile and the horizontal dotted line shows the observed terminal velocity $v_{\infty}$. The vertical gray region shows what is beyond the star's surface.}
    \label{fig:velo_prof}
\end{figure}

In Sect. \ref{sec:energetics}, we assumed that the entire envelope is ejected with a single asymptotic velocity. This remains a crude approximation, as it neglects the detailed stellar density structure. A more realistic treatment would adopt a velocity–stratified ejecta \citep[e.g.][]{2020MNRAS.497.3166I,2017MNRAS.471.3200M}. Instead of assuming every layer leaves at the observed velocity $v_{\infty}$, we also considered an alternative model in which we parametrized the ejecta velocity as a radius-weighted power law, scaled to the local escape velocity:
\begin{equation}
v_{\rm ej}(m) = \xi(m) v_{\rm esc}(m) \quad \text{with} \quad
\xi(m) =\left(\dfrac{r(m)}{R_0}\right)^{b} \dfrac{v_\infty}{v_{\rm esc}^{\rm surf}} 
\end{equation}
where $R_{0}$ is the radial coordinate at the surface of the star and  $v_{\rm esc}^{\rm surf}$ is the escape velocity at $R_{0}$. The parameter $b$ is the radial power-law index controlling the degree of velocity stratification in the ejecta, i.e. how the outflow velocity changes from the inner layers to the surface. The corresponding velocity profiles are illustrated in Fig.~\ref{fig:velo_prof}. Values of $b>0.5$ produce steeper velocity gradients, with outer layers moving significantly faster than deeper ones. The profile $\xi(m)$ suppresses the contribution of deeper layers, reflecting their larger binding energy: inner material receives a smaller effective boost and thus attains lower terminal velocities, while the outer layers are accelerated to higher velocities. $\xi(m)$ has been designed to reproduce the observed velocity at infinity $v_{\infty}$ for the outer layers. The escape velocity is given by:
\begin{equation}
v_{\rm esc}(m) =
\begin{cases}
\sqrt{\dfrac{2 G(m+M_{a})}{r(m)}} & \text{for } m > m_{\mathrm{He}} , \\
+\infty & \text{for } m < m_{\mathrm{He}} 
\end{cases}
\end{equation}
where $m$ is the mass coordinate, and $m_{He}$ is the helium-core mass of the donor. The local escape velocity of the stripped donor increases steeply at the helium–core boundary\footnote{For AT2021biy, adopting $M_{d}=22M_{\odot}, q=5$ and $a=700R_{\odot}$, the helium-core mass $m_{\mathrm{He}}$ roughly corresponds to 60$\%$ of the donor mass at the onset of the RLOF.}, reflecting the transition to a compact core that is effectively non-ejectable in the standard CE energy formalism. 

The bulk kinetic energy of the unbound ejecta in Eq. \ref{eq:bulk_kinetic} can therefore be expressed by taking the contributions from each individual layers independently. This implies that outer layers are typically faster while deeper layers are slower. This yields the more realistic, velocity-stratified description of the ejecta:

\begin{equation}
\label{eq:velo_prof}
E_{ej}(M_{\text{env}})= \frac{1}{2}\int_{M_{d,ce} - M_{\text{env}}}^{M_{d,ce}}v_{ej}^2(m)dm\,.
\end{equation}

Adopting such a prescription can lead to substantially different ejecta masses compared to the assumption of a single asymptotic velocity for each layer. In the specific case of a homologous velocity profile ($b=1$), the inferred ejecta masses can be up to a factor of two higher (see Fig. \ref{fig:candidates_vejm}). Consequently, lower ejection velocities for the inner layers prevents overestimating the total kinetic energy of the ejecta. 

 
\section{Summary and conclusion}
\label{sec:conclusion}

In this work, we proposed the first study combining observations from before, during, and after the outburst of two LRNe and their progenitor systems. By combining MESA binary-evolution models with light-curve based models and a dust analysis, we constrained the progenitor properties and estimated the ejecta and dust masses of two extragalactic LRN events, AT2021biy and AT2021blu. 

More precisely, we proposed a refined methodology that builds upon the approach of \cite{Blagorodnova2021} by explicitly applying mass-transfer instability criteria within MESA binary models to identify systems that undergo unstable evolution and produce an LRN outburst. Given the large size of the parameter space for possible progenitor systems, our approach cannot identify all viable binary configurations. Instead, our primary goal has been to delineate the region of parameter space that can plausibly host the progenitors of these LRNe, rather than to provide a complete census of all possible binary systems. If we identify the onset of instability with the point of OLOF in the models, we find the following constraints on the progenitors: $M_{d}=18-23M_{\odot}$, $q\gtrsim3$ and $a=300-700R_{\odot}$ for AT2021biy and $M_{d}\sim14 \pm 0.5 M_{\odot}$, $q\gtrsim10$ and $a=200-300R_{\odot}$ for AT2021blu. These estimates are in agreement with the values reported in \cite{Cai2022AA} and \cite{2023A&A...671A.158P}, respectively.

Building on this progenitor modeling, we used MESA predictions to construct a simple energy-budget method based on energy conservation. Under the idealized assumption of perfect conversion of the released orbital energy into unbinding the envelope, we estimated for each progenitor candidate, the envelope mass that can be ejected during the CE episode. Accounting for both the lower and upper limits provided by our method, We found the ranges $M_{ej}=0.03-7.82 M_{\odot}$ for AT2021biy and $M_{ej}=0.02-3.8M_{\odot}$ for AT2021blu, depending on the mass ratio considered. We compared these estimates with predictions from light-curve models and found better consistency in the intermediate mass-ratio regime ($3\lesssim q\lesssim10$ for AT2021biy and $5<q\lesssim15$ for AT2021blu). 

Both the energetic approach and the light-curve-based models are highly sensitive to the assumed ejecta velocity, implying that uncertainties in this parameter propagate significantly into the inferred masses. We further showed that adopting a more realistic ejecta-velocity prescription, in which the outflow speed increases with radius, lowers the bulk kinetic energy of the ejecta. In the specific case of a homologous velocity profile, this translates into higher estimates for the lower limit of the ejected envelope mass.

Finally, we used NEOWISE detections to model the evolution of the dust mass at multiple epochs for both events. The inferred values are 1–5 orders of magnitude lower than the estimated ejected-envelope masses, which is naturally explained if only a small fraction of the expelled gas condenses into dust. The dust-mass evolution is consistent with shock-powered interaction with pre-existing circumstellar material. That interpretation is further supported by our MESA-based upper limits of substantial mass loss during the binary interaction, with $M_{loss}=0.2-0.5 M_{\odot}$ for AT2021biy and $M_{loss}=0.05-0.3 M_{\odot}$ for AT2021blu. The presence of shocks also suggests that some fraction of the dust may be destroyed over time.

In future work, we plan to refine our MESA calculations by relaxing the point-mass approximation for the companion and treating its internal structure and evolution self-consistently. This will provide a more realistic description of the companion–donor interaction, including possible accretion, stripping, and feedback on the mass-transfer history. We also intend to make use of the CE modules available in MESA to derive constraints from observations obtained closer to the outburst. More generally, we aim to move beyond the assumption of spherical symmetry by incorporating guidance from three-dimensional hydrodynamic simulations calibrated on the transient light curves, in order to refine our estimates of the ejected mass. Finally, late-time infrared observations would offer direct constraints on the dust mass and thereby improve the limits on the minimum mass ratio.


  
%


\input{acknowledgements}

%
%

\bibliographystyle{aa}
\bibliography{ref}

\begin{appendix} 
\nolinenumbers
\section{Physical ingredients}
\label{app:phys_ingredients}

In all our models, convective regions are determined with the classical \cite{1947ApJ...105..305L} criterion and the convection mechanism itself is modeled on the basis of the mixing-length theory introduced by \citet{1958ZA.....46..108B} and described in \cite{1968pss..book.....C}. The mixing length parameter $\alpha=1.5$ defines that the average distance travelled by a gas parcel before being completely dissolved is one and a half times the height of the local pressure scale. We assumed a highly efficient mixing process \citep{1983A&A...126..207L} at the boundary with a semi-convection parameter of $\alpha_{sc}=100$ in agreement with recent studies on massive stars in the SMC \citep{2019A&A...625A.132S}. We adopted these parameters calibrated on SMC massive stars as a proxy, even though they were derived at a lower metallicity than our progenitors. Unfortunately, we did not find a more suitable constraint on $\alpha_{sc}$ for our specific case. Note that the choice of $\alpha_{sc}$ can impact the post-MS expansion phase as we can read in Appendix B of \cite{2020A&A...638A..55K} where assuming solar metallicity, an inefficient semiconvection leads to the formation of separate convective zones. Thermohaline mixing in our models is based on the theory developed by \cite{1980A&A....91..175K}, and we adopted a dimensionless efficiency coefficient $\alpha_{th}=1$. This mixing mechanism accounts for composition inversions that can arise, for instance, during mass accretion or shell burning phases.

The hydrogen-burning core is allowed to overshoot into the radiative envelope during the main sequence. Following the step overshooting mechanism calibrated by \cite{2011A&A...530A.115B} for B-type giants ($\sim$15 $M_{\odot}$) in the Large Magellanic Cloud (LMC), we adopted an overshooting parameter $f_{ov}=0.345$, corresponding to an overshooting length of approximately 0.345 pressure scale heights.We assumed the same overshooting parameter for the helium-burning core as for the hydrogen-burning core and, following \citet{2019A&A...625A.132S}, neglected overshooting in shell-burning regions as well as in convective zones within stellar envelopes. All other extra mixing processes, including rotationally induced mixing mechanisms such as Eddington–Sweet circulations, have also been neglected in these models.

The stellar winds have been modeled following \cite{2011A&A...530A.115B} and using the metallicity dependence of mass-loss rates $\dot{M}\propto (Z/Z_{\odot})^{0.85}$ for hydrogen-rich massive stars \citep{2001A&A...369..574V}. Depending on the star properties, i.e., its hydrogen abundance $X_{s}$ and effective temperature $T_{\mathrm{eff}}$, different models were used. We adopted the wind mass-loss rates from \citep{2000A&A...362..295V,2001A&A...369..574V} in the case $(X_{s} > 0.7,\ T_{\mathrm{eff}} > 25\mathrm{kK})$, rescaled values from \cite{1995A&A...299..151H} for hydrogen-poor stars $X_{s} < 0.4$, and interpolation between the two for intermediate hydrogen abundances $0.4 < X_{s} < 0.7$. Finally, for $T_{\mathrm{eff}} < 25\mathrm{kK}$, we used the maximum of the rates calculated by \cite{2000A&A...362..295V} and \cite{1990A&A...231..134N}.

The nuclear reactions are assumed to follow the standard networks provided by MESA with \texttt{basic.net} for H and He burning, and \texttt{co\_burn.net} for C and O burning. MESA also employs a blended EOS constructed from several prescriptions and studies \citep[e.g.,][]{1995ApJS...99..713S,2010CoPP...50...82P}. Similarly, the works of \citet{1996ApJ...464..943I} and \citet{2005ApJ...623..585F} were used to establish the radiative opacities.

We model rotational transport in the diffusion approximation following the MESA implementation based on \citet{2000ApJ...528..368H,2005ApJ...626..350H}. We include rotationally induced chemical mixing from Eddington–Sweet circulation, the Goldreich–Schubert–Fricke instability, and secular shear, with an overall efficiency factor $f_{c}=1/30$. Dynamical shear and Solberg–Høiland mixing are disabled. Angular-momentum transport includes the Spruit–Tayler mechanism (without associated chemical mixing). The inhibiting effect of mean-molecular-weight gradients is reduced using the standard MESA choice 
$f_{\mu}=0.05$.

\section{Light curves based models}
\label{app:LC_based_models}

The first model we considered is that of \citet{2009ApJ...703.2205K}, originally developed for Type IIP supernovae. Their framework combines 1D hydrodynamical simulations of exploding red supergiant stars with detailed radiative-transfer calculations, where the resulting light curves and spectra are primarily governed by hydrogen recombination, radiative diffusion, and radioactive decay. From these simulations, the authors derived analytic scaling relations linking observable properties to the explosion energy, ejecta mass, and progenitor radius. Applied to our transients, this formalism predicts ejecta masses differing by nearly three orders of magnitude, with ($M_{ej}=0.5125 M_{\odot}$) for AT2021biy and ($M_{ej}=0.0009 M_{\odot}$) for AT2021blu.

In \citet{2017ApJ...835..282M}, the authors propose to explain the light curves of the M31LRN 2015 stellar-merger transient by distinguishing two components. The first would result from fast ejecta produced by shocks as the companion plunges into the envelope of the primary star, while the second would be associated with a more substantial mass loss occurring during the deeper inspiral phase. Applying this model to our events, we found a difference of two orders of magnitudes between AT2021biy ($M_{ej}=3.92 M_{\odot}$) and AT2021blu ($M_{ej}=0.06 M_{\odot}$).

The model presented by \citet{1993ApJ...414..712P} provides analytic scaling relations to describe the plateau phase of Type II-P supernova light curves. It mostly assumes a homologously expanding envelope, constant opacity, and recombination-driven photon release, showing that the plateau luminosity and duration primarily depend on the explosion energy, progenitor radius, ejecta mass, and envelope opacity. Later, \citet{2017MNRAS.472..224S} found a good agreement between this model with modified absolute coefficients and 1D flux-limited diffusion simulations of Type II-P supernova light curves. Here, the estimated ejecta mass of AT2021blu ($M_{ej}=0.28 M_{\odot}$) differs by nearly one order of magnitude from the unreliable estimate obtained for AT2021biy ($M_{ej}=19.82 M_{\odot}$).

In \citet{2017MNRAS.471.3200M}, the authors attribute the observed light curves to a two-stage process. Before the merger, a slow equatorial mass is lost through the outer Lagrange point and form an outflow orbiting the system. This slow moving material later interacts with the dynamically rapidly ejected shell. The first peak in luminosity arises from the cooling of this fast ejecta, while the second is driven by shocks generated as it collides with the pre-existing equatorial material. We assumed that the shocked shell moves at 5\% of the ejecta velocity, and that the pre-existing wind has a radial velocity equal to 10\% of the ejecta velocity. We further assumed that this pre-existing wind has been formed through previous mass loss during 1000 days and occupies 30\% of the solid angle. Of all the models presented in Table \ref{tab:ejecta_mass}, this one shows the smallest discrepancy in the estimated ejecta mass between AT2021biy ($M_{ej}=6.25 M_{\odot}$) and AT2021blu ($M_{ej}=0.55 M_{\odot}$).

As shown in \citet{2022ApJ...938....5M}, the double-peaked structure commonly observed in LRNe can be simply described by a stratification of the ejecta, in the case of spherical symmetry, where each layer would possess different velocities and internal energies. The duration of the second peak or plateau is attributed to the energy released by hydrogen recombination, and it depends on the timescale for this process to be fully realized. The direct comparison of AT2021biy and AT2021blu light curves clearly reveals a difference in the duration of the second peak. According to equation (16) in \citet{2022ApJ...938....5M}, this difference can be explained by both the average velocity of the ejecta and their mass, as well as the density of each layer. This observation is further supported by the fact that AT2021blu ($6.5\times10^{40}$ erg.s$^{-1}$) is a less energetic event \citep[see][]{Cai2022AA,2023A&A...671A.158P} than AT2021biy ($1.6\times10^{41}$ erg.s$^{-1}$). This implies lower velocities, ejecta mass and thereby recombination timescale. The predictions from this model show a difference of more than one order of magnitude between AT2021biy ($M_{ej}=9.58 M_{\odot}$) and AT2021blu ($M_{ej}=0.29 M_{\odot}$). 

\section{Alternative energy-based CE algorithm}
\label{new_algo}
We propose here a shell-based prescription to model the inspiral of the companion inside the donor envelope, allowing for non-local and delayed envelope ejection.

Once the companion enters the donor envelope, we assumed that the motion remains quasi-circular, and the instantaneous separation can be approximated by the semi-major axis $a$.
The inspiral starts at the donor surface $a_0 = R_{\rm d}$ (an additional component, defined below, accounts for the orbital decay from $a_{\rm OLOF}$) and proceeds inward through a discretized sequence of separations $a_i$.

At each step $i \to i+1$, we compute the following contributions.
The orbital energy released between two successive separations is
\begin{equation}
\Delta E_{\rm orb,i}
=
-\frac{G\,M_{\rm enc}(a_i)\,M_{a}}{2a_i}
+\frac{G\,M_{\rm enc}(a_{i+1})\,M_{a}}{2a_{i+1}},
\end{equation}
where $M_{\rm enc}(a_i)$ is the donor mass enclosed within radius $a_i$ and $M_{a}$ is the companion mass. 
The binding-energy cost of unbinding the shell between $a_i$ and $a_{i+1}$ is written as
\begin{equation}
\Delta E_{\rm bind,i}
=
\int_{M_{\rm enc}(a_{i+1})}^{M_{\rm enc}(a_i)}
\left(\frac{Gm}{r(m)} - u(m)\right)\,{\rm d}m,
\end{equation}
where $r(m)$ is the radius coordinate of mass shell $m$, and $u(m)$ is the specific internal energy.
We also include the bulk kinetic-energy requirement of the corresponding shell,
\begin{equation}
\Delta E_{\rm ej,i}
=
\frac{1}{2}\,v_{\rm ej}^2(m_i)\,\Delta m_i,
\end{equation}
where $\Delta m_i = m(a_i) - m(a_{i+1})$ is the shell mass and $v_{\rm ej}(m_i)$ is the prescribed homologous expansion (see Eq. \ref{eq:velo_prof} with $b=1$) evaluated at that shell.

At each step, the available energy is defined as
\begin{equation}
E_{\rm avail,i} = E_{\rm res,i} + \Delta E_{\rm orb,i},
\end{equation}
where $E_{\rm res,i}$ is the energy reservoir that stores any leftover energy from previous steps. $E_{\rm res,0}$ represents an initial orbital energy injection that completes the energy budget by accounting for the orbital decay from $a_{OLOF}$ down to the donor's surface.
To account for delayed (non-local) ejection, we introduce an energy debt, $E_{\rm debt,i}$, which stores any unfulfilled energy requirements from earlier steps.
The total energy required at step $i$ is then
\begin{equation}
E_{\rm due,i}
=
E_{\rm debt,i}
+\Delta E_{\rm bind,i}
+\Delta E_{\rm ej,i}.
\end{equation}
The ejected fraction is computed as
\begin{equation}
f_i = \min\!\left(\frac{E_{\rm avail,i}}{E_{\rm due,i}},\,1\right).
\end{equation}
The reservoir and debt are updated according to
\begin{equation}
    E_{\rm res,i+1} = E_{\rm avail,i} - f_i\,E_{\rm due,i} \quad \text{and} \quad
E_{\rm debt,i+1} = (1-f_i)\,E_{\rm due,i}.
\end{equation}
The iteration is terminated when the companion’s orbit reaches the final separation $a^{*}_{f}$ as defined in Eq. \ref{eq:final_orbit}.

Fig.~\ref{new_algo} illustrates this alternative method for estimating the ejecta mass for an AT2021biy progenitor with $M_{d}=22M_\odot$, $q=5$ and $a=700R_{\odot}$. More specifically, for each shell we show the effective ejected mass and the corresponding fraction of the shell that is unbound. We found that, down to $2M_{\odot}$ below the stellar surface, the available energy exceeds the total energy required at each iteration step, allowing each layer to be fully ejected. At greater depths, however, the binding energy begins to dominate the energy budget, and the resulting cumulative energy deficit rapidly reduces the fraction of each shell that can be unbound. A non zero fraction of each shell can be ejected down to the merger zone corresponding to the condition given in Eq. \ref{eq:final_orbit}. We also examined the ratio between the angular momentum carried away by the ejecta and the total orbital angular momentum change during inspiral, $\Delta J_{\rm loss,i}/ \Delta J_{\rm orb,i}$. We found that this ratio increases progressively down to $2M_{\odot}$ below the stellar surface, beyond which it drops sharply as the ejected mass decreases. The ratio remains predominantly below unity, indicating that the angular momentum carried away by the ejecta is insufficient to account for the total orbital angular momentum change. This suggests that additional mechanisms, such as hydrodynamical drag or gravitational torques, must contribute to the orbital evolution.

\begin{figure}[h!]
    \centering
    \includegraphics[width=0.5\textwidth]{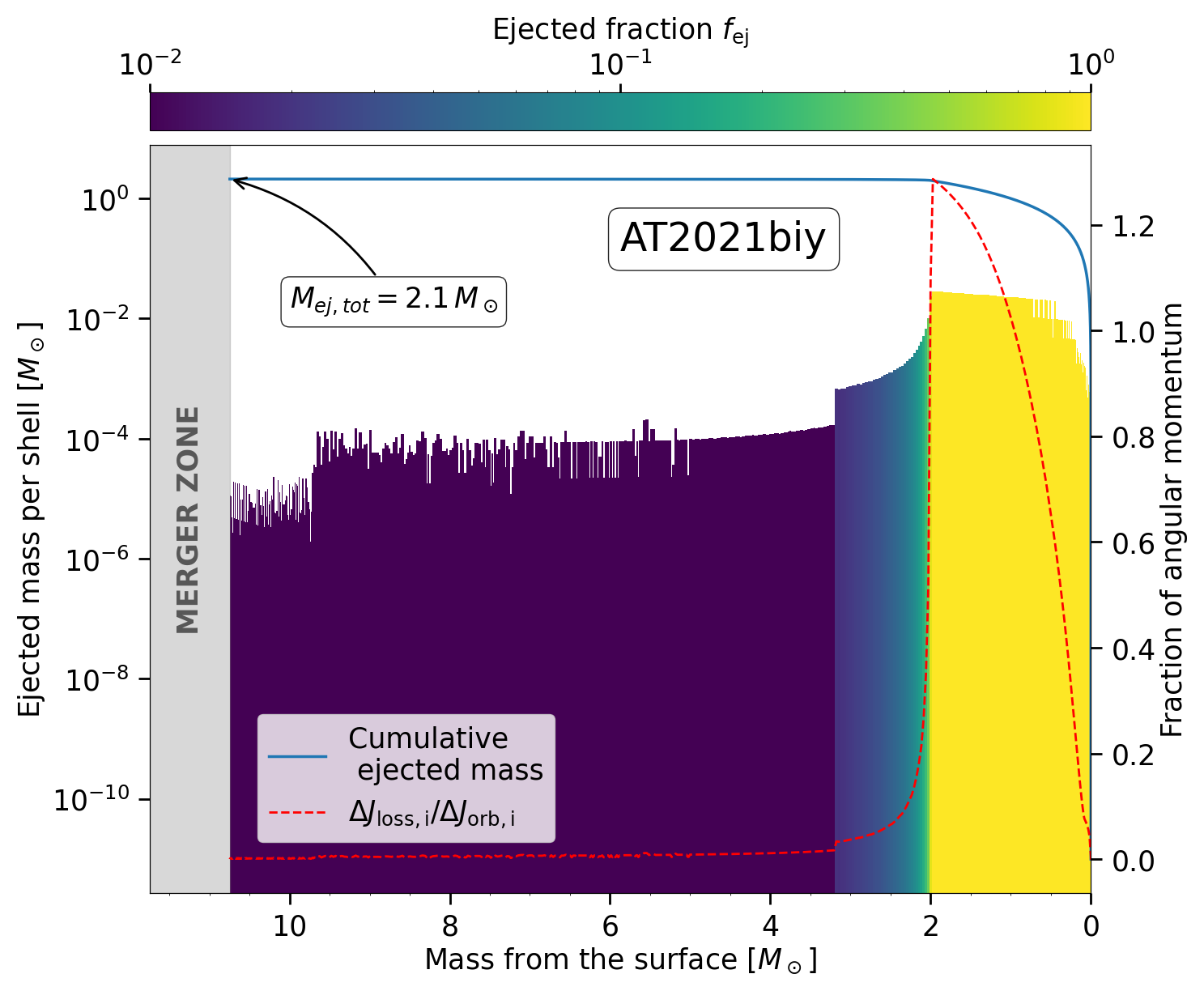}
    \caption{Ejected mass per shell computed with the alternative algorithm for a AT2021biy progenitor with $M_{d}=22M_\odot$, $q=5$ and $a=700R_{\odot}$. We also displayed in red dashed line, the ratio between angular momentum lost through ejecta and the total orbital angular momentum change during inspiral.}
    \label{fig:AT2021biy_newalgo}
\end{figure}

\section{Additional figures}

\begin{figure*}[h!]
    \centering
    \includegraphics[width=\textwidth]{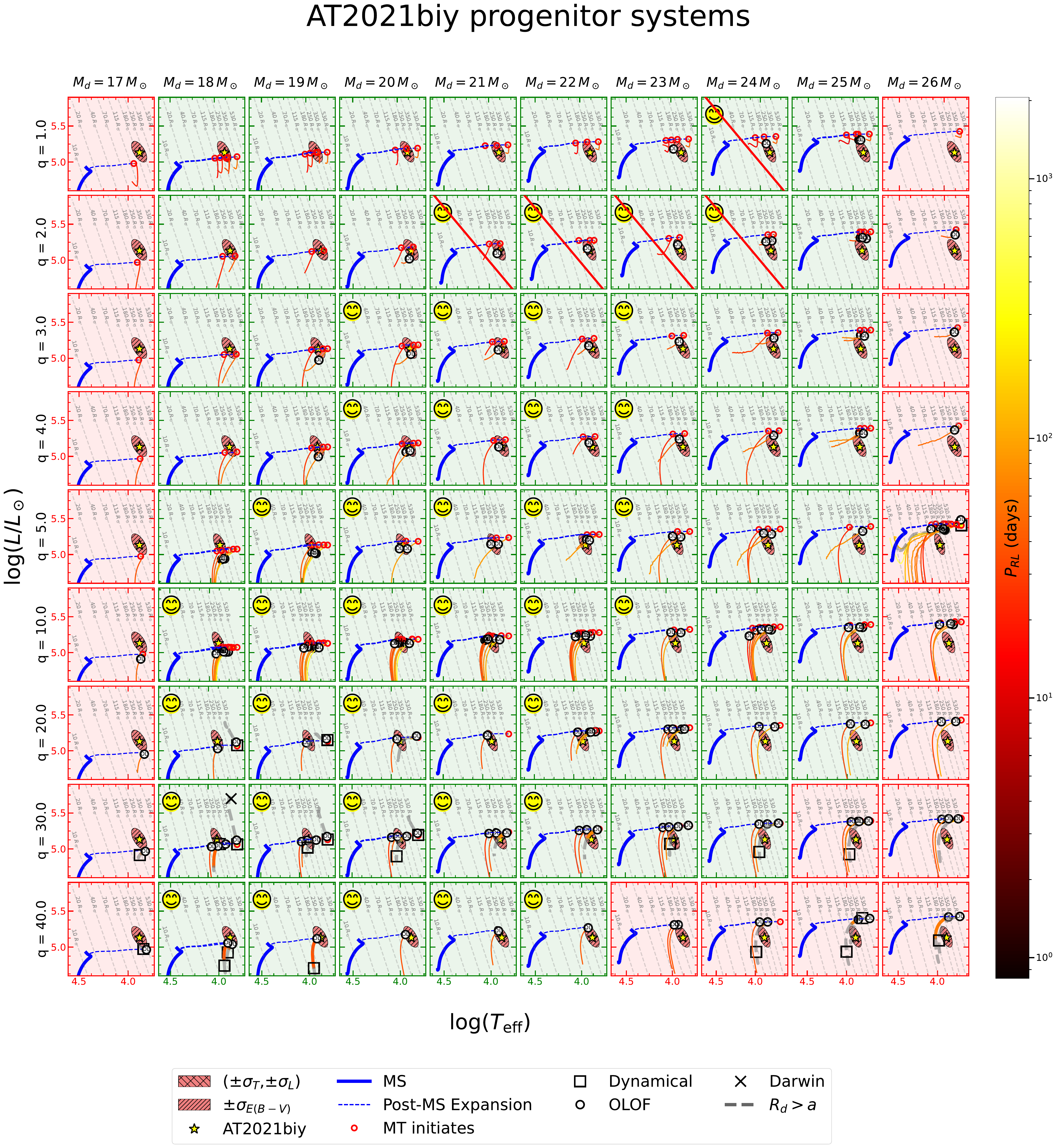}
    \caption{Grid of stellar evolutionary tracks for various configurations of donor star mass, mass ratio, and initial binary separation for AT2021biy. Progenitors consistent with the HST observations not only during the dynamical mass-transfer phase are highlighted in green, while the remaining systems are shown in red. Binary systems that satisfy our instability criterion based on the onset of OLOF and match the observational detection are marked with a yellow smiley. The mass transfer phase starts at the red circle. The color bar indicates the binary system's period at the onset of Roche Lobe overflow $P_{\rm RL}$. A progenitor candidate for which the onset of the OLOF arises within the HST region, but the mass transfer rate never enters a runaway, is crossed out in red.}
    \label{fig:AT2021biy_grid}
\end{figure*}

\begin{figure*}[h!]
    \centering
    \includegraphics[width=\textwidth]{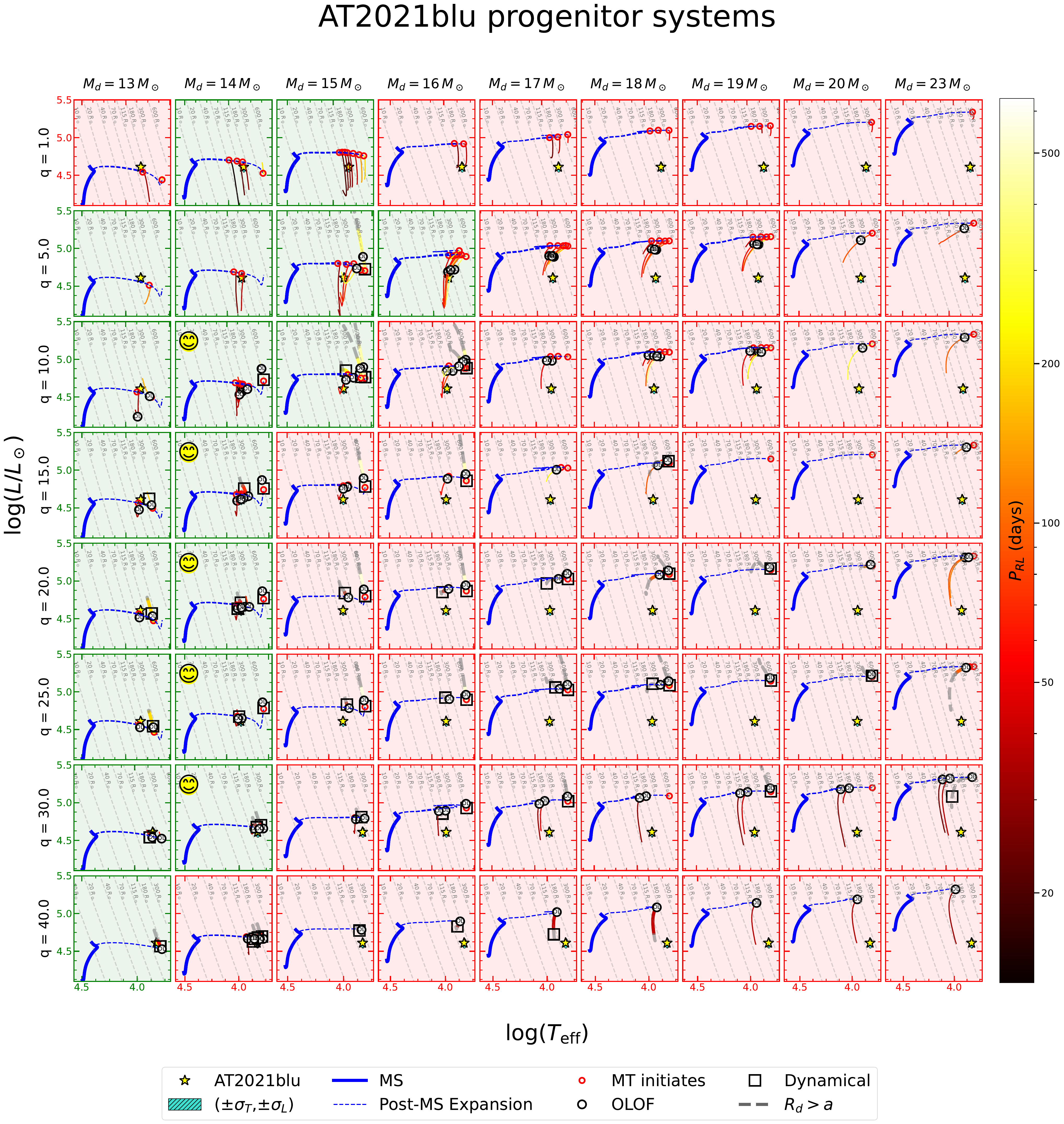}
    \caption{Same as Fig. \ref{fig:AT2021biy_grid}, showing the grid of stellar evolutionary tracks for AT2021blu.}
    \label{fig:AT2021blu_grid}
\end{figure*}

\begin{figure*}[h!]
    \centering
    \includegraphics[width=\linewidth]{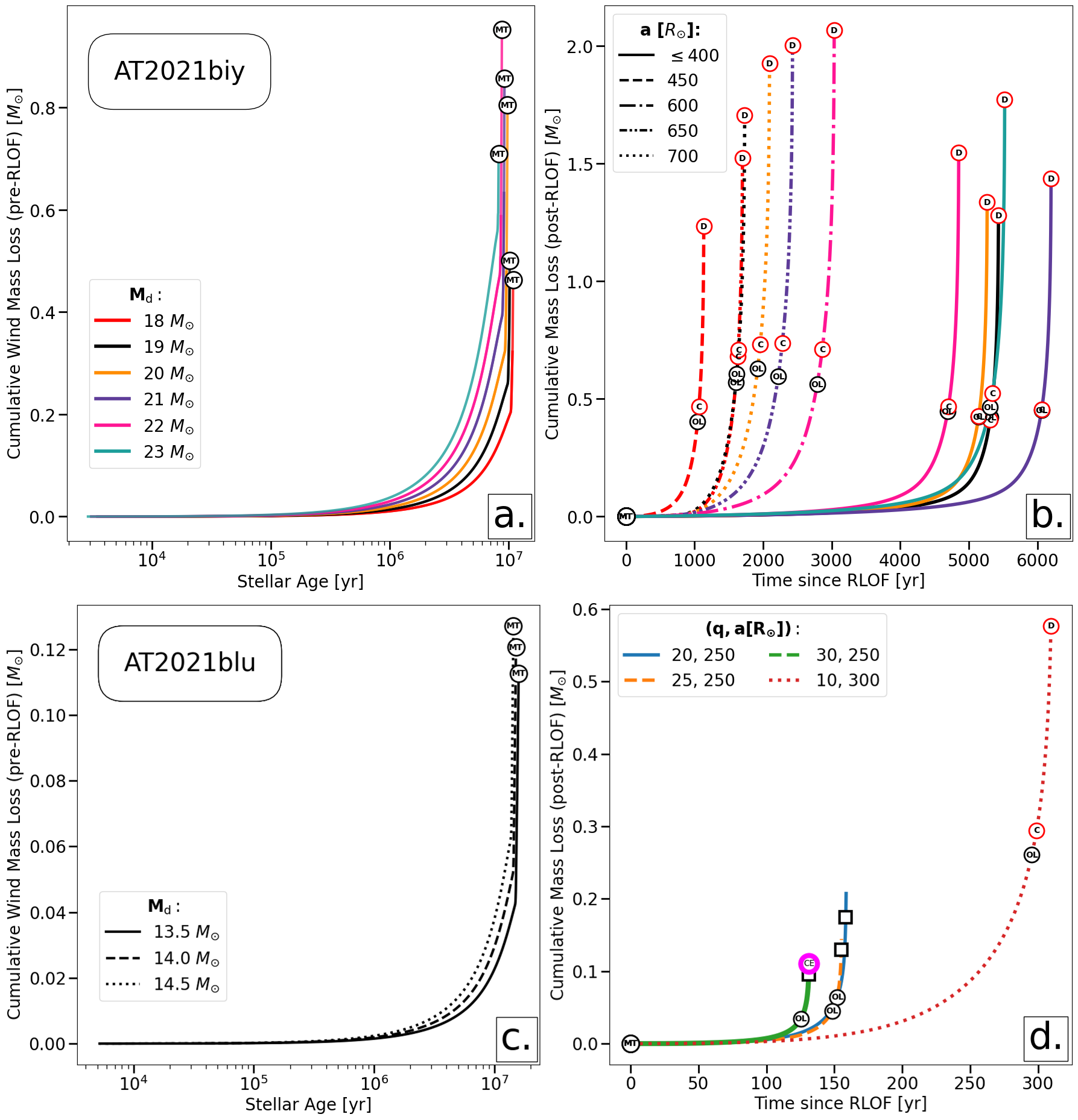}
    \caption{The cumulative mass loss before and after RLOF for the progenitors displayed in Fig. \ref{fig:candidates}. We displayed by \textbf{C}) the moment of the maximum overfilling factor and \textbf{D)} the moment of max $\dot{M}_{d}$. In the case of AT2021biy (top panels), all the displayed progenitors correspond to a mass ratio $q=10$. We displayed the position of the different instabilities and evolution steps mentioned before: 'MT' for the onset of mass-transfer through RLOF, 'OL' for the OLOF, the dynamical instability is shown by a square and the onset of CE by a magenta circle.  The time axis of the right panel starts when the mass-transfer rate exceeds the wind mass-loss rate for the first time.}
    \label{fig:candidates_ML}
\end{figure*}

\begin{figure}[h!]
    \centering
    \includegraphics[width=0.5\textwidth]{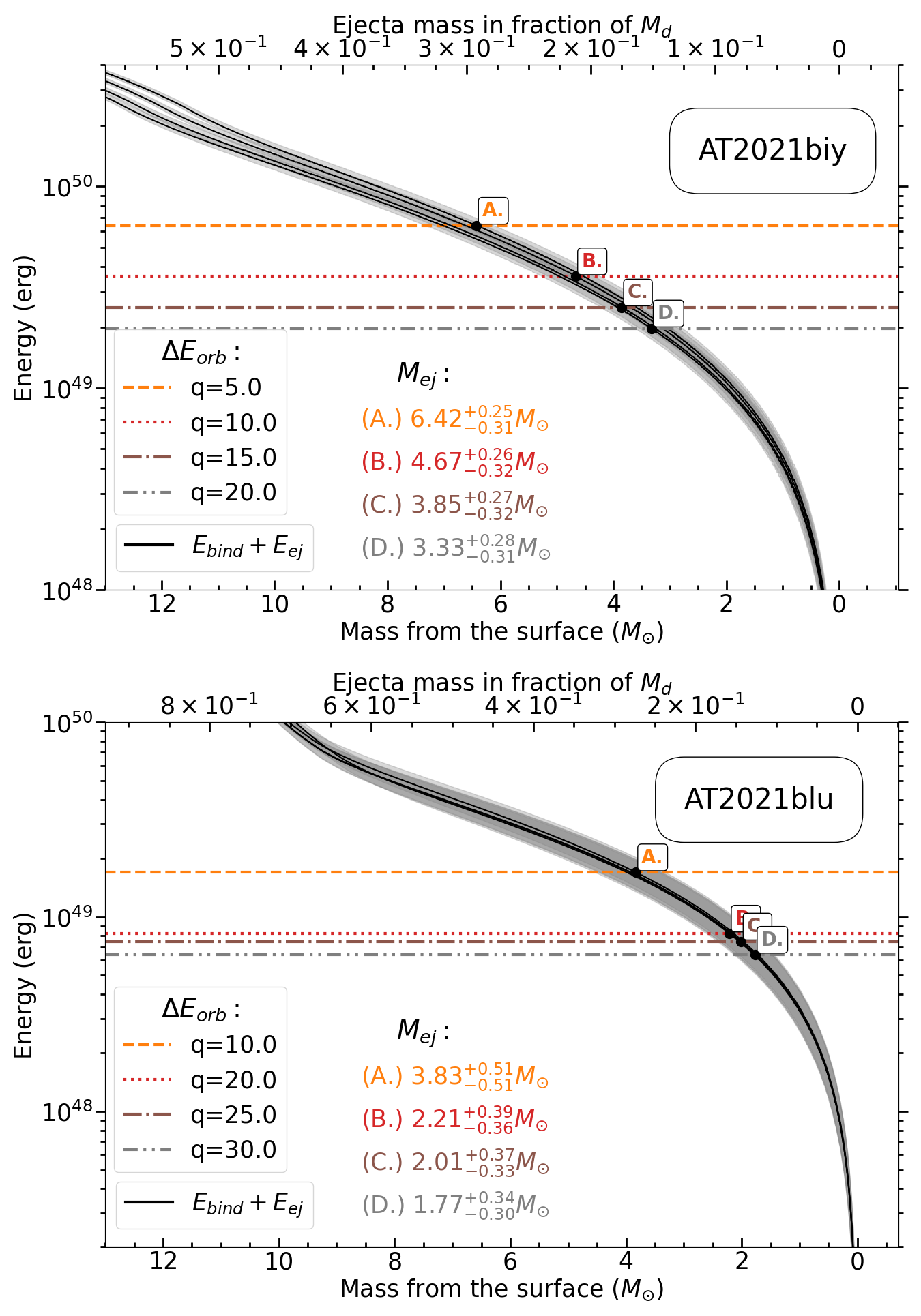}
    \caption{The ejecta mass $M_{ej}$ inferred from the energy-budget at the moment of the OLOF for each mass ratio (intersections A,B,C and D). We assumed that the companion inspirals until it merges with the helium core of the donor. The grey thickness around the envelope binding energy and the bulk kinetic energy of the ejecta $E_{\rm bind}+E_{\rm ej}$ (black curve) represents the uncertainties on the observed ejecta velocity ($\pm 90\mathrm{km\,s^{-1}}$ for AT2021biy and $\pm 15\mathrm{km\,s^{-1}}$ for AT2021blu). We used the same MESA binary models than presented in Fig. \ref{fig:AT2021biy_energy}.}
    \label{fig:upperlimit}
\end{figure}

\begin{figure}[h!]
    \centering
    \includegraphics[width=\linewidth]{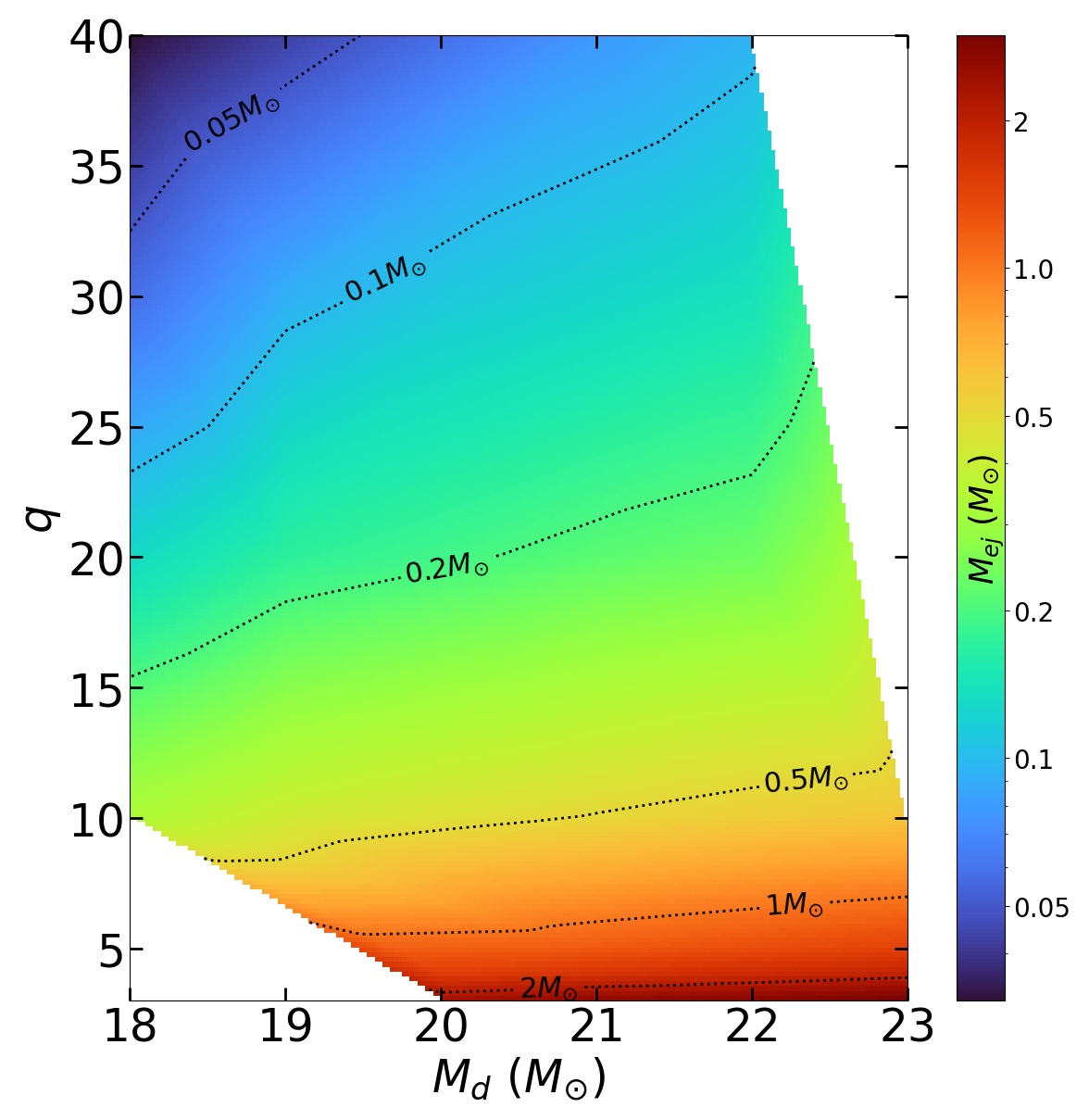}
    \caption{Interpolated estimates for the lower limit of the ejected envelope mass, based on the values computed with the method described in Sect. \ref{sec:energetics} for AT2021biy. The colour scale indicates the ejected mass, which were derived for $M_{d}=18-23M_\odot$ and $q=3-40$.}
    \label{fig:AT2021biy_Mej_grid}
\end{figure}

\begin{figure}[h!]
    \centering
    \includegraphics[width=0.5\textwidth]{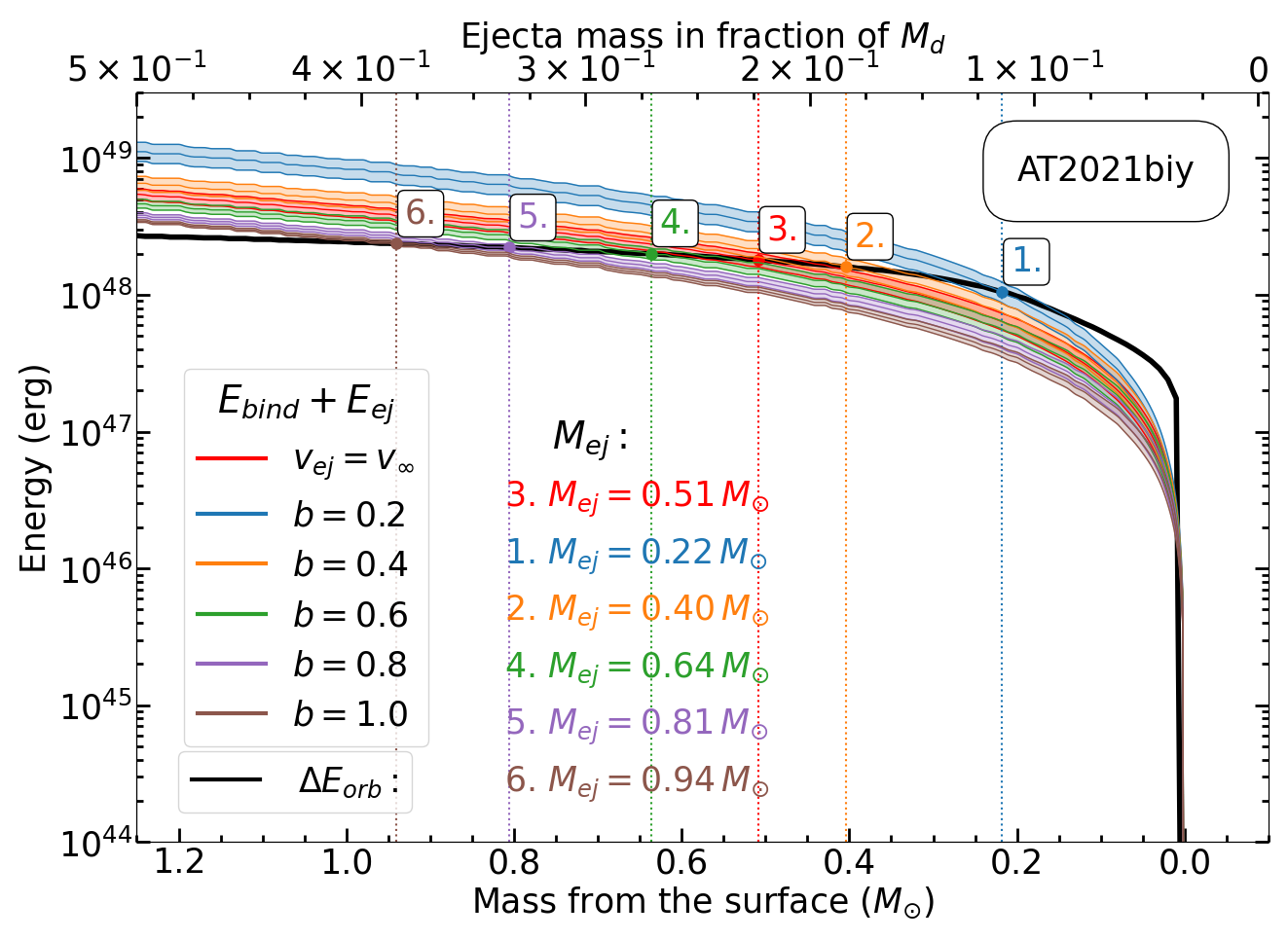}
    \caption{Lower-limit estimates of the ejected envelope mass for different velocity profiles. Using MESA predictions at the onset of OLOF for an AT2021biy progenitor with $M_{d}=22 M_{\odot}$ and $q=10$, we compare the orbital energy released with the envelope binding energy and the bulk kinetic energy of the ejecta associated with each velocity profile. This figure extends the method illustrated in Fig.~\ref{fig:AT2021biy_energy} by allowing for velocity-stratified ejecta and shows how this assumption affects the lower-limit estimate of the ejecta mass.}
    \label{fig:candidates_vejm}
\end{figure}

\end{appendix}
\end{document}

%% file: acknowledgements.tex
\begin{acknowledgements}
The authors are grateful to Jacob Jencson and Guido de Marchi for their advice on photometry, particularly regarding the use of DOLPHOT, and to Ondřej Pejcha for his guidance in interpreting the outputs of 3D hydrodynamics simulations. M.~W., N.~B., and M.~A.~G.-M. acknowledge to be funded by the European Union (ERC, CET-3PO, 101042610). Views and opinions expressed are however those of the author(s) only and do not necessarily reflect those of the European Union or the European Research Council Executive Agency. Neither the European Union nor the granting authority can be held responsible for them. The authors finally acknowledge financial support from the Maria de Maetzu institute grant CEX2024-001451-M funded by MICIU/AEI/10.13039/501100011033.
\end{acknowledgements}